\documentclass{article}
\pdfoutput=1
\usepackage{arxiv}
\setcitestyle{authoryear,open={(},close={)}}
\renewcommand{\cite}{\citep}

\usepackage{hyperref}
\usepackage{authblk}
\usepackage{url}
\usepackage{multirow}
\usepackage{graphicx}
\usepackage{booktabs}
\usepackage{xcolor}
\usepackage{array}
\usepackage[utf8]{inputenc} 
\usepackage{wrapfig}
\usepackage[T1]{fontenc}    
\usepackage{hyperref}       
\usepackage{url}            
\usepackage{booktabs}       
\usepackage{amsfonts}       
\usepackage{nicefrac}       
\usepackage{microtype}      
\usepackage{xcolor}         
\usepackage{amsmath}
\usepackage{amssymb}
\usepackage{mathtools}
\usepackage{amsthm}
\usepackage{graphicx, subcaption}
\usepackage{tablefootnote}

\usepackage[capitalize,noabbrev]{cleveref}
\usepackage{xspace}
\usepackage[table]{xcolor}
\usepackage{algorithm}
\usepackage{algpseudocode}
\usepackage{booktabs}
\usepackage{makecell}
\usepackage{pifont}
\newcommand{\cmark}{\textcolor{blue!50!black}{\ding{51}}}
\newcommand{\xmark}{\textcolor{red}{\ding{55}}}

\hypersetup{
    colorlinks,
    citecolor=[rgb]{0.00, 0.45, 0.70}, %
    linkcolor=[rgb]{0.01, 0.62, 0.45}, %
    urlcolor=[rgb]{0.68, 0.45, 0.63} %
}
\definecolor{green1}{rgb}{0.01, 0.62, 0.45}
\definecolor{blue1}{rgb}{0.00, 0.45, 0.70}
\definecolor{blue2}{rgb}{0.612, 0.8, 0.902}
\definecolor{purple1}{rgb}{0.68, 0.45, 0.63}
\definecolor{red1}{rgb}{1.0, 0.1, 0.3}
\definecolor{red2}{rgb}{1.0, 0.77, 0.80}

\usepackage{todonotes}
\usepackage{enumitem}
\usepackage{amsmath}
\usepackage{amsthm}
\usepackage{amssymb}
\usepackage[bbgreekl]{mathbbol}

\DeclareMathSymbol\drule  \mathord{bbold}{"01}

\newcounter{packednmbr}

\newenvironment{packeditemize}{\begin{list}{$\bullet$}{\setlength{\itemsep}{0.5pt}\addtolength{\labelwidth}{-4pt}\setlength{\leftmargin}{2ex}\setlength{\listparindent}{\parindent}\setlength{\parsep}{1pt}\setlength{\topsep}{2pt}}}{\end{list}}

\usepackage{bbm}

\usepackage{wrapfig}
\usepackage{graphicx}
\usepackage{caption}
\usepackage{titletoc}
\usepackage{placeins}
\usepackage{graphicx}
\usepackage{placeins}
\usepackage[most]{tcolorbox} %
\usepackage{pdflscape}
\usepackage{algorithm}
\usepackage{tabularx}
\usepackage{siunitx}
\usepackage{soul}

\newcommand{\tnote}[1]{{\color{orange} [TM:#1]}}

\renewrobustcmd{\bfseries}{\fontseries{b}\selectfont}
\renewrobustcmd{\boldmath}{}
\newrobustcmd{\B}{\bfseries}
\usepackage[utf8]{inputenc} 
\usepackage[T1]{fontenc}    
\usepackage{hyperref}       
\usepackage{url}            
\usepackage{booktabs}       
\usepackage{amsfonts}       
\usepackage{nicefrac}       
\usepackage{microtype}      
\usepackage{xcolor}         
\usepackage{amsmath}
\usepackage{amssymb}
\usepackage{pifont}
\usepackage{xspace}
\usepackage{graphicx}
\usepackage{wrapfig}
\usepackage{algpseudocode}
\usepackage{color}
\usepackage{tabu}
\usepackage{diagbox}
\usepackage{float}
\usepackage{placeins}
\usepackage{multirow}
\usepackage{tikz}
\usepackage{graphicx}
\usepackage{subcaption}
\usepackage{colortbl}
\usepackage{threeparttable}
\definecolor{stepfill}{RGB}{20,20,20}   
\definecolor{cmtgray}{gray}{0.5}

\newcommand{\stepbadge}[1]{%
  \tikz[baseline=-0.55ex]{%
    \node[circle, fill=stepfill, text=white, inner sep=0pt,
          minimum size=1.15em, font=\sffamily\scriptsize\bfseries] {#1};}%
}

\newcommand{\sctx}{\mathcal{C}}   
\newcommand{\tque}{\mathcal{T}}   

\newcommand{\stage}[2][]{%
  \hfill
  \ifx\relax#1\relax\else
    \textcolor{cmtgray}{\itshape #1}\hspace{0.45em}%
  \fi
  \stepbadge{#2}%
}

\algrenewcommand\algorithmicrequire{\textbf{Input:}}
\algrenewcommand\algorithmicensure{\textbf{Output:}}

\definecolor{plugcolor}{RGB}{255,243,205} 

\definecolor{rowgray}{gray}{0.93}
\NewDocumentCommand{\yuzhen}{ mO{} }{\textcolor{red}{\textsuperscript{\textit{yuzhen}}\textsf{\textbf{\small[#1]}}}}
\newcommand{\toolName}{\textsc{DeLM}\xspace}

\definecolor{plugbg}{HTML}{DCEEFB}
\definecolor{plugfg}{HTML}{124780}

\newtcolorbox{qualitativeBox}{
  colback=gray!10, %
  colframe=gray!20, %
  rounded corners, %
  boxrule=0.5pt, %
  left=10pt, %
  right=10pt, %
  top=5pt, %
  bottom=5pt, %
  boxsep=5pt, %
}
\newtcolorbox{ourMethodBox}{
  colback=blue1!10, %
  colframe=gray!20, %
  rounded corners, %
  boxrule=0.5pt, %
  left=10pt, %
  right=10pt, %
  top=5pt, %
  bottom=5pt, %
  boxsep=5pt, %
}
\newtcolorbox{theirMethodBox}{
  colback=red1!10, %
  colframe=gray!20, %
  rounded corners, %
  boxrule=0.5pt, %
  left=10pt, %
  right=10pt, %
  top=5pt, %
  bottom=5pt, %
  boxsep=5pt, %
}

\definecolor{gisttext}{HTML}{24292E}
\definecolor{gistbg}{HTML}{F4F6F8}
\definecolor{gistframe}{HTML}{CDD4DC}
\definecolor{gistfact}{HTML}{157F5F}
\definecolor{gistfail}{HTML}{C0392B}
\definecolor{gisttid}{HTML}{5A67B8}
\lstdefinestyle{gist}{%
  basicstyle=\footnotesize\ttfamily\color{gisttext},
  backgroundcolor=\color{gistbg},
  frame=single,
  framerule=0.7pt,
  rulecolor=\color{gistframe},
  framesep=7pt,
  framexleftmargin=7pt,
  xleftmargin=9pt,
  xrightmargin=5pt,
  breaklines=true,
  breakatwhitespace=false,
  columns=fullflexible,
  keepspaces=true,
  aboveskip=0.9\baselineskip,
  belowskip=0.7\baselineskip,
  emph={[1]FACT},emphstyle={[1]\bfseries\color{gistfact}},
  emph={[2]FAIL},emphstyle={[2]\bfseries\color{gistfail}},
  emph={[3]t0,t1,t2,t3,t4,t5},emphstyle={[3]\color{gisttid}},
}

\newtcolorbox{keyidea}{
  enhanced, breakable,
  colback=blue!8!white, colframe=blue!55!black,
  boxrule=0.6pt, arc=2pt,
  left=4pt, right=4pt, top=4pt, bottom=4pt,
  before skip=6pt, after skip=6pt   
}

\begin{document}

\title{Decentralized Multi-Agent Systems with Shared Context}
\author{
    {Yuzhen Mao\thanks{Correspondence to yuzhenm@stanford.edu}, ~
    Jerry Gu, ~
    Aadi Chauhan, ~
    Qizheng Zhang, ~
    Hangoo Kang, ~
    Azalia Mirhoseini}\linebreak\linebreak
    {Stanford University}\linebreak
}


\date{}

\maketitle

\begin{abstract}
Multi-agent systems (MAS) can scale large language model agents on long-horizon tasks by running them in parallel, yet existing designs waste much of this parallelism in \emph{bubbles}: agent time spent waiting on others or redoing a peer's work. These bubbles stem from how agents communicate. Independent agents share nothing and rediscover what their peers have already found; peer-communicating agents wait at synchronous rounds; and under centralized orchestration, the main agent blocks on its sub-agents while progress is relayed. We propose Decentralized Language Models (\toolName{}), a MAS framework on top of existing agent harnesses that squeezes out these bubbles by replacing the main agent with a shared context and a task queue. Agents asynchronously claim tasks, publish findings as soon as they are available, and build on or correct one another's progress, with every peer's status visible to all. On long-horizon tasks from Terminal-Bench 4.0 and DeepSWE v1.1, and on SWE-bench Verified, \toolName{} is both more accurate and faster than Codex, Claude Code, their native subagents, and AOrchestra in every
setting, improving accuracy by up to 17.5 points over the strongest
baseline and running up to 2.49$\times$ faster than the harness it
builds on. On ProgramBench, where agents rebuild programs from scratch, \toolName{} makes faster progress than Claude Code and finishes a 120-minute budget up to 19.9 points higher in test pass rate.
The code is available on our project website at~\url{https://yuzhenmao.github.io/DeLM/}.
\end{abstract}

\begin{figure}[H]
    \centering
    \includegraphics[width=1\linewidth]{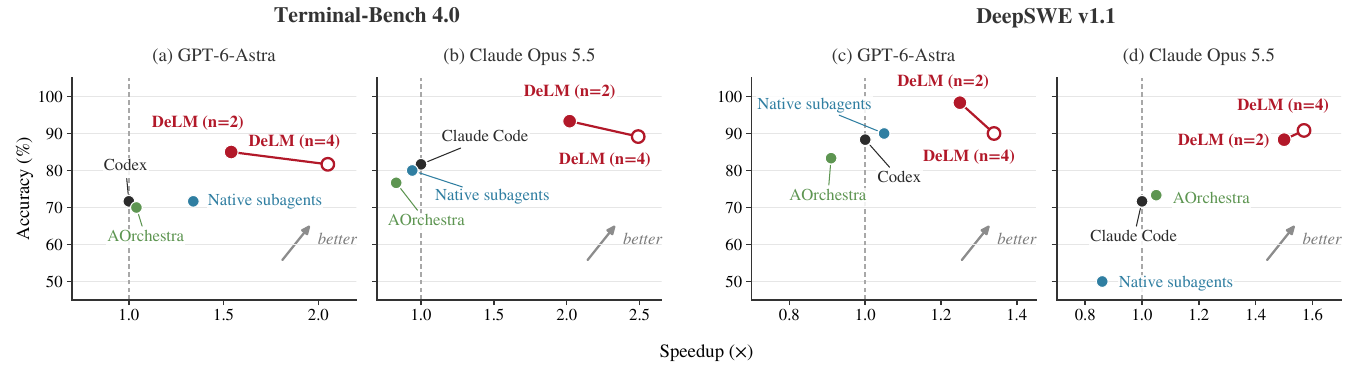}
    \caption{\textbf{DeLM is both faster and more accurate than existing multi-agent coding systems.}
    Averaged over the two models, DeLM ($n{=}2$) is 1.78$\times$ faster and 12.5 points more accurate
    than Codex/Claude Code (dashed line, 1$\times$) on Terminal-Bench 4.0, and 1.38$\times$ faster and
    13.3 points more accurate on DeepSWE v1.1. Each benchmark uses 10 long-horizon tasks (\S\ref{sec:experiments}).}
    \label{fig:teaser}
\end{figure}
\section{Introduction}
\label{sec:intro}

Multi-agent systems (MAS) offer a natural way to scale large language model reasoning at test time: instead of solving a complex task in a single trajectory, they decompose it into subtasks, dispatch agents in parallel, and aggregate their progress~\citep{hong2023metagpt, feng2026agentswing, ruan2026aorchestra, zhang2025recursive, openai_codex_2025, anthropic_claude_code_2026, team2026kimi}.
This paradigm is increasingly important in agentic coding, both in repository-level software engineering (e.g., SWE-bench~\citep{jimenez2024swebench}, DeepSWE~\citep{datacurve2026deepswev11}), where agents can investigate different files, root causes, or fixes concurrently, and in interactive terminal tasks (e.g., Terminal-Bench~\citep{merrill2026terminal}), where agents can explore alternative solution paths while sharing what they learn about the environment.
In both settings, the value of MAS lies not in issuing more model calls, but in converting additional test-time compute into useful parallel progress, reducing wall-clock time and improving success rates without a proportional increase in cost.

\begin{figure}[t]
    \centering
    \includegraphics[width=0.72\linewidth]{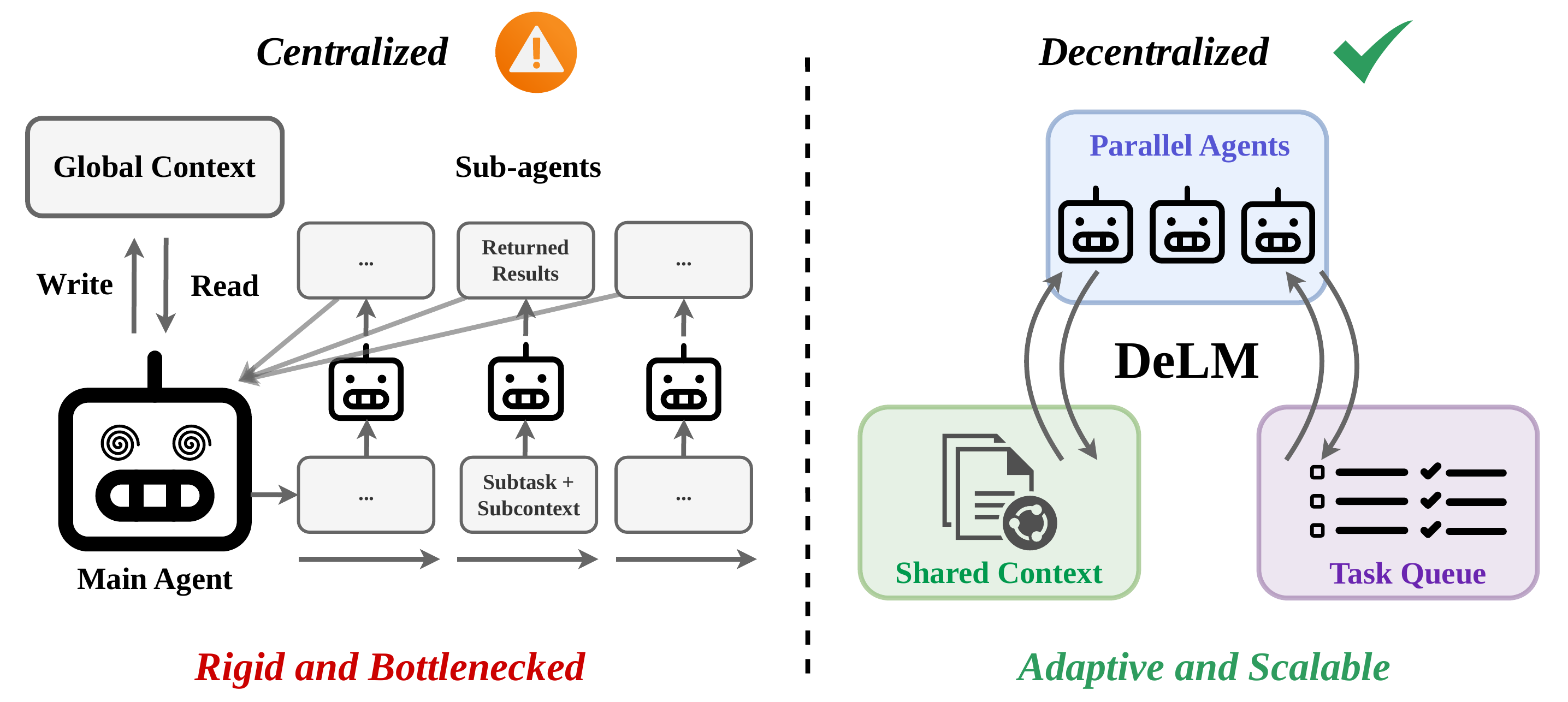}
    \caption{
\textbf{Centralized vs. decentralized multi-agent systems.} Centralized MAS relies on a main agent to assign subcontexts, spawn sub-agents, and integrate their results, so progress is shared mainly through the main agent and sub-agents remain unaware of one another's work. In contrast, \toolName{} decentralizes coordination through parallel agents, a shared context, and a task queue: agents asynchronously claim ready tasks, publish findings incrementally to a shared context, and build on or correct each other's progress without a main agent in the loop.
}
    \label{fig:first}
\end{figure}

Realizing this potential requires keeping every agent productive. Borrowing a term from pipeline parallelism~\citep{huang2019gpipe}, we call agent time that does not advance the solution a \emph{bubble}: time an agent spends waiting on others, or redoing work a peer has already done. Bubbles waste compute and stretch wall-clock time, and each of the three dominant families of MAS creates them in its own way.

\textbf{Independent agents} run in parallel and select among their outputs, as in self-consistency~\citep{wang2022self}, repeated sampling~\citep{brown2024large}, and best-of-$n$ sampling~\citep{cobbe2021training,lightman2024let}. They never wait, but they share nothing while they run, so each agent must rediscover the same faults, fixes, and dead ends on its own. Their bubbles are \emph{redundant work}: adding agents adds compute, but much of it repeats exploration another agent has already done.

\textbf{Peer-communicating agents} exchange responses over synchronous rounds, as in multi-agent debate~\citep{du2023improving,liang2023encouraging,chen2024reconcile}. Seeing peers' answers lets agents correct one another, but each round is a barrier: its duration is set by the slowest agent, so agents that finish early sit idle. These \emph{barrier waits} are the straggler problem familiar from bulk-synchronous parallel computing~\citep{valiant1990bridging,dean2008mapreduce}. Communication cost also grows with the number of agents: in the fully connected setting~\citep{du2023improving}, each agent ingests all $n-1$ peer responses every round, for $n(n-1)$ messages per round, so each agent's context fills with peer responses it must read and reconcile on its own.

\begin{figure}[t]
\centering
\includegraphics[width=0.9\linewidth]{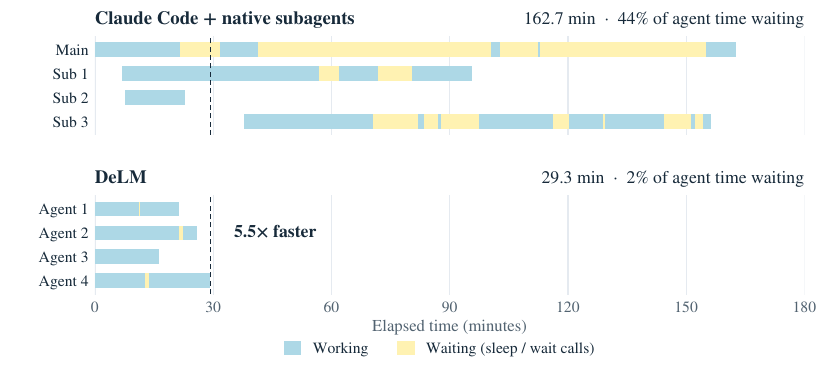}
\caption{\textbf{Centralized orchestration creates bubbles; \toolName{} squeezes them out.}
Execution timelines on Terminal-Bench 4.0's \texttt{retro-console-soc} task with Claude Opus 5.5 and four agents.
With Claude Code's centralized native subagents, the main agent spends 73\% of its time waiting on delegated work (yellow), and 44\% of all agent time is spent waiting.
\toolName{} agents build on one another's intermediate results, spend 2\% of their time waiting, and finish 5.5$\times$ sooner.}
\label{fig:bubbles}
\end{figure}

\textbf{Centralized orchestration} underlies structured reasoning methods~\citep{yao2023tree,besta2024graph,ning2024skeleton}, SPRINT~\citep{biju2026sprint}, the native subagents of Claude Code~\citep{anthropic_claude_code_2026} and Codex~\citep{openai_codex_2025}, AOrchestra~\citep{ruan2026aorchestra}, and blackboard-style systems with a control agent~\citep{han2025exploring,salemi2025llm}. A main agent decomposes the task, dispatches subtasks, waits for their outputs, and integrates the results (Figure~\ref{fig:first}, left). This parallelizes sub-agent execution but not the coordination around it, creating \emph{relay bubbles}: the main agent idles while delegated work runs, and sub-agents, unaware of one another, cannot build on each other's progress until the main agent relays it. On Terminal-Bench 4.0's \texttt{retro-console-soc} task (Figure~\ref{fig:bubbles}), Claude Code's main agent spends 73\% of its time waiting on its subagents. Relaying can also lose information. Because progress lives in the main agent's context, details can be dropped in summarization, and correcting an error that has already propagated requires another round of delegation.

\begin{table}
\centering
\small
\caption{Collaboration and coordination mechanisms in the compared multi-agent systems. A checkmark means the system provides the mechanism while a cross means that it does not.}
\setlength{\tabcolsep}{6pt}
\resizebox{\textwidth}{!}{%
\begin{tabular}{l*{6}{c}}
\toprule
& \multicolumn{3}{c}{General collaboration capabilities}
& \multicolumn{3}{c}{Explicit coordination mechanisms} \\
\cmidrule(lr){2-4}
\cmidrule(lr){5-7}
System
& \makecell{Parallel subtask\\execution}
& \makecell{Cross-agent\\work reuse}
& \makecell{Feedback \&\\correction}
& \makecell{Intermediate-results\\sharing}
& \makecell{Peer-status\\awareness}
& \makecell{Stateful\\shared context}  \\
\midrule
Isolated agents
& \xmark & \xmark & \xmark
& \xmark & \xmark & \xmark \\

AOrchestra
& \xmark & \cmark & \cmark
& \xmark & \xmark & \xmark \\

Codex
& \cmark & \cmark & \cmark
& \cmark & \xmark & \xmark \\

Claude Code
& \cmark & \cmark & \cmark
& \cmark & \xmark & \xmark \\

\rowcolor{rowgray}\textbf{\toolName{}}
& \cmark & \cmark & \cmark
& \cmark & \cmark & \cmark \\
\bottomrule
\end{tabular}%
}
\label{tab:agent-comparison}
\end{table}
We propose \textbf{Decentralized Language Models (\toolName{})}, a MAS framework that squeezes out these bubbles by replacing the main agent with a shared problem state (Figure~\ref{fig:first}, right). \toolName{} consists of parallel agents, a shared context, and a task queue, and each component targets one kind of bubble. Agents claim tasks from the queue asynchronously, so no agent waits at a round barrier or for a main agent to assign work. They publish compact findings to the shared context as soon as results are available, so peers build on intermediate progress instead of waiting for a relay. The queue and shared context also show what each agent is working on and which approaches have failed, so agents avoid redoing a peer's work. Finally, corrections are appended as new entries rather than relayed through a main agent, so a fix reaches every agent without another round of delegation. Table~\ref{tab:agent-comparison} contrasts these mechanisms with those of the systems we compare against. On Terminal-Bench 4.0's \texttt{retro-console-soc} task (Figure~\ref{fig:bubbles}), four \toolName{} agents spend only 2\% of their time waiting and finish 5.5$\times$ sooner than Claude Code with native subagents.

\begin{keyidea}
\textbf{Key idea.} Existing MAS waste parallelism in bubbles because of how their agents communicate: independent agents redo one another's work, peer-communicating agents wait at synchronous rounds, and centralized systems wait on a main agent to relay progress. \toolName{} squeezes out these bubbles by letting agents coordinate asynchronously through a shared context and a task queue, with no main agent in the loop.
\end{keyidea}

We evaluate \toolName{} on four agentic coding benchmarks: Terminal-Bench 4.0~\citep{merrill2026terminal}, where agents share discovered environment state and failed attempts in live terminals; DeepSWE v1.1~\citep{datacurve2026deepswev11}, where agents reuse each other's debugging progress on long-horizon software engineering tasks; SWE-bench Verified~\citep{jimenez2024swebench}, where agents explore alternative root causes and fixes for real GitHub issues; and ProgramBench~\citep{yang2026programbenchlanguagemodelsrebuild}, where agents rebuild entire programs from scratch. We compare against Codex and Claude Code, with and without native subagents, AOrchestra~\citep{ruan2026aorchestra}, and, on SWE-bench Verified, mini-SWE-agent~\citep{yang2024sweagent}. Our key findings are:
\begin{packeditemize}
    \item \textbf{\toolName{} is both faster and more accurate on long-horizon coding.} Across two frontier models, \toolName{} reaches up to 93.33\% on 10 long-horizon tasks of Terminal-Bench 4.0 and 98.33\% on 10 long-horizon tasks of DeepSWE v1.1, improving over the strongest baseline by up to 17.5 points while running up to 2.49$\times$ faster than the harness it builds on (Figure~\ref{fig:teaser}, \S\ref{sec:results-tb40}, \S\ref{sec:results-deepswe11}).
    \item \textbf{The gains also hold with a lightweight model on a standard benchmark.} On SWE-bench Verified with Gemini 3 Flash, \toolName{} reaches 66.08\%, 10.8 points above the strongest baseline, while running 2.12$\times$ faster than Claude Code at half of AOrchestra's cost per submission (\S\ref{sec:swe}).
    \item \textbf{\toolName{} provides an effective way for scaling agents, enabling more agents to make more progress under a fixed time budget.} On ProgramBench, where agents rebuild entire programs from scratch, \toolName{} ends a 120-minute budget up to 19.91 points above Claude Code, and matches Claude Code's final pass rate in half the time (\S\ref{sec:program-bench}).
    \item \textbf{The gains come from coordination, not parallelism alone.} Removing any single coordination mechanism brings two agents back to roughly single-agent speed (\S\ref{sec:ablation}).
\end{packeditemize}
\section{Decentralized Language Models (\toolName{})}
\label{sec:method}
\begin{figure}[t]
    \centering
    \includegraphics[width=0.7\linewidth]{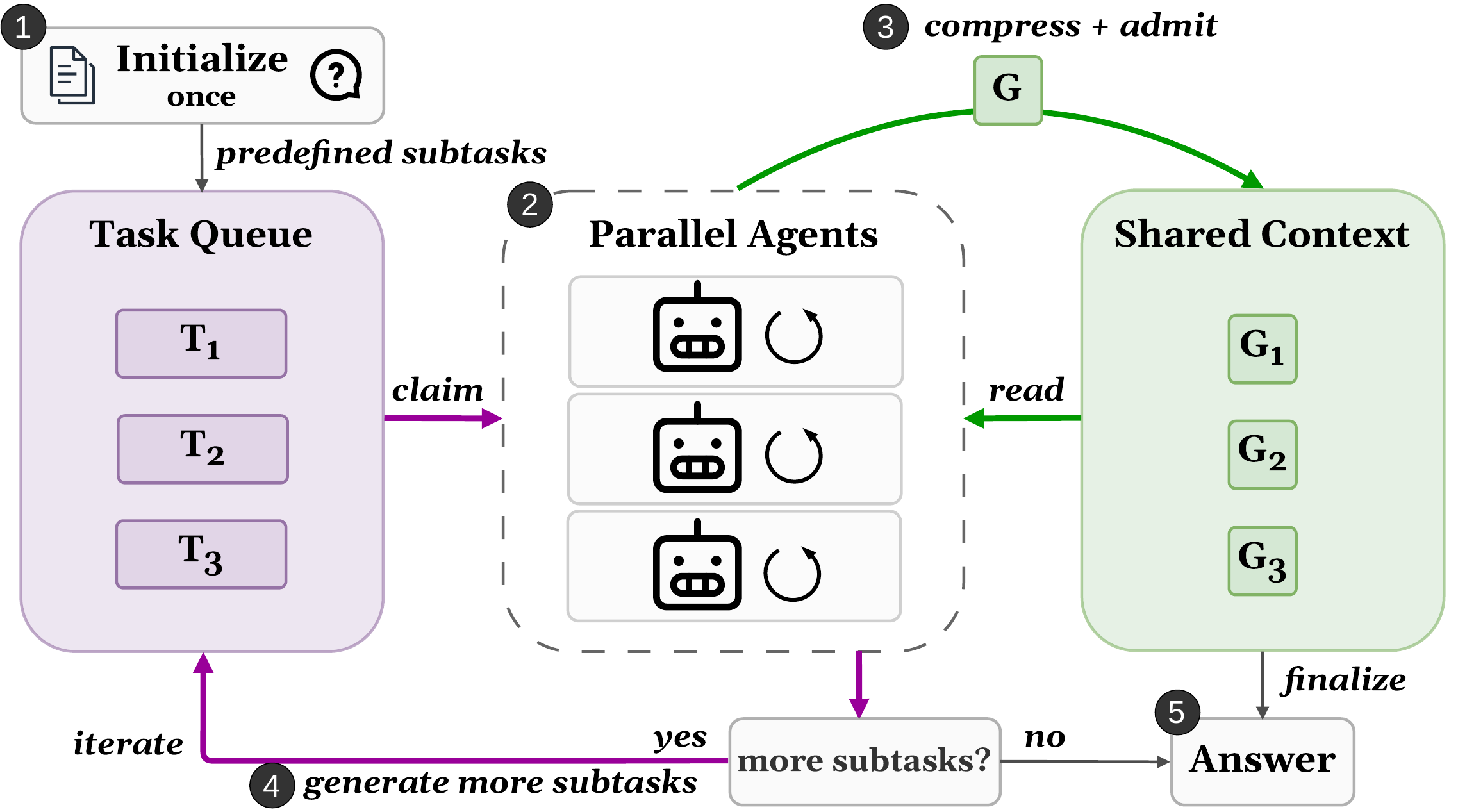}
    \caption{\textbf{Overview of \toolName{}.} The shared task queue is initialized with a set of predefined, general preparation subtasks. Parallel agents asynchronously claim subtasks $\{T_i\}$, read the shared context, and implement $\{T_i\}$ locally. Each agent's update is compressed and appended to the shared context as a gist $G_i$, making reusable progress visible to all agents. There is no central coordinator or reviewer: any agent may add required work at any time, and each agent submits its own complete solution once its evidence covers all task requirements.
}
    \label{fig:pipeline}
\end{figure}

Given a task $D$, \toolName{} runs $n$ agents concurrently, each with its own persistent session and workspace. Agents coordinate through two shared structures, a shared context $\sctx$ and a task queue $\tque$, and each agent produces its own complete solution.

Figure~\ref{fig:pipeline} and Algorithm~\ref{alg:dlm} summarize five operations\mbox{: \stage{1}}~initialize the task queue\mbox{, \stage{2}}~execute subtasks in parallel\mbox{, \stage{3}}~summarize and publish intermediate results\mbox{, \stage{4}}~update the task queue as work progresses\mbox{, and \stage{5}}~finalize each agent's solution. These operations proceed asynchronously rather than in lockstep: while one agent publishes a result, others keep working and can reuse it as soon as it appears.

\subsection{Shared Context}\label{sec:shared-context}
The shared context $\sctx$ is an append-only log of reusable findings and contributions from all agents.
Agents write to it through compact entries $G_i$ of a declared type: a \textsc{fact} records a finding supported by observations, a \textsc{fail} records an approach or hypothesis that was contradicted, and a completion summary records the result of a finished subtask. \textsc{fact} and \textsc{fail} entries carry explicit scope and evidence fields; agents are instructed to keep entries brief and to state unfinished parts and known limitations, so that other agents can judge whether an entry applies to their own work. Any entry may attach files, which peers can inspect as diffs or merge into their own workspaces with conflict protection. Agents publish an entry as soon as a result is usable and then continue with their own work.

\paragraph{Summaries with files by reference.}
An entry can attach files, which are stored separately from its summary and snapshotted at publication time. Agents inspect these files with \texttt{expand}, which shows them as diffs against the original task files, and copy selected files into their own workspace with \texttt{apply}. \texttt{apply} refuses to overwrite local changes it cannot account for, leaving reconciliation to the receiving agent. A read of $\sctx$ shows only what is new since that agent's last read, namely new entries and changed work status, while older entries remain retrievable on demand. This design resembles demand paging in an operating system: agents follow progress through compact summaries and fetch detailed content from backing storage only when they need it (Appendix~\ref{sec:unfold}).


\paragraph{Distributed verification and correction.}
Verification is left to the agents rather than to a central checker: each agent derives the checks for its own solution from the original task rather than from subtask summaries, reuses applicable evidence from its peers without repeating it, and directs additional checks at requirements and component interactions that existing evidence does not cover. Because entries are immutable, an agent that finds a defect in a shared result does not edit it. Instead, it appends a \textsc{fail} entry stating the conditions under which the defect occurs, and a correction follows as a new entry or a repair subtask. A \textsc{fail} stays open until it is addressed, and before submitting, each agent reads the shared context and resolves any published failure relevant to its own solution.


\subsection{Task Queue}
The task queue $\mathcal{T}$ divides the work among agents. Each item consists of a subject and a free-text detail, in which agents are instructed to state the intended output, its available inputs, the interface it shares with other work, and the earliest partial result it depends on. Agents are instructed to create items only for required implementation, repair of a demonstrated defect, or a concrete open question the solution depends on. At startup, the first agent reserves its own scope and publishes complementary items for its peers, or states the dependency that prevents a split; if it ends before this handoff, the next running agent takes over. After that, any agent can add items or claim open ones at any time. There is no privileged planner and no plan-closure step.

\paragraph{Claims and status.}
The queue enforces exclusive claims: the first claim wins, and each agent holds at most one unfinished claim. Every item is marked open, claimed, or done, together with its owner, so each agent can see who is responsible for what and how much remains. Each read also shows every agent's session state (starting, running, or ended), tracked by the framework rather than reported by the agents. Agents use this state to choose complementary work and to handle dependencies. If a dependency's owner is starting or running, an agent continues with independent work and waits only when nothing else remains, using a bounded wait that returns as soon as a peer publishes or ends. If the owner of an item has ended without completing it, this does not block progress: an agent that needs the result builds it in its own workspace, reusing anything the owner already shared, instead of waiting.

\paragraph{When agents publish and read.}
Apart from session state, which the framework tracks automatically, all
content in the shared context and task queue comes from the agents
themselves: the framework never publishes on an agent's behalf. Agents
decide when to publish and read under the instructions in
Appendix~\ref{sec:prompts}. An agent publishes when a result can unblock
a peer, without needing to run new checks first, and avoids routine
status updates and repeated evidence. It reads the shared context at
startup, before starting a new subproblem or diagnostic hypothesis,
before final integration, and before submitting. How often an agent publishes and reads is therefore driven mainly by its own progress rather than by the number of peers.
\section{Experiments}
\label{sec:experiments}


\paragraph{Benchmarks.}
We evaluate \toolName{} on four agentic coding benchmarks.
(1) \textbf{Terminal-Bench 4.0}~\citep{merrill2026terminal}. The benchmark consists of tasks that agents solve in live terminal environments, such as configuring systems, installing dependencies, and debugging failures. The tasks require long sequences of shell commands and agents benefit from sharing discovered environment state and failed attempts.
(2) \textbf{DeepSWE v1.1}~\citep{datacurve2026deepswev11}. The benchmark contains long-horizon software-engineering tasks. Agents must localize faults, understand interdependent modules, and debug across large codebases. It tests whether agents can reuse each other's localization and debugging progress instead of repeating it.
(3) \textbf{SWE-bench Verified}~\citep{jimenez2024swebench}. The benchmark is built from real-world GitHub issues, with human-verified task specifications and tests. Agents inspect the repository, identify the relevant files, implement a fix, and validate it through tests. 
(4) ProgramBench~\citep{yang2026programbenchlanguagemodelsrebuild}. Each task gives the agent a compiled executable and its documentation, and asks it to re-implement the program from scratch. Submissions are scored by hidden behavioral tests that compare the candidate against the original program. We test ProgramBench on two hard tasks, \texttt{ctags} and \texttt{pandoc}, and track the test pass rate over a fixed 120-minute budget. 

Because \toolName{} targets long-horizon tasks where coordination matters, we evaluate on a subset of long-running tasks from Terminal-Bench 4.0 and DeepSWE v1.1. We select tasks based on the wall-clock latency of Codex with GPT-6-Astra: for Terminal-Bench 4.0, we select 10 tasks with latency between 15 minutes and 1 hour; for DeepSWE v1.1, we select 10 tasks with latency of at least 10 minutes. The full list of selected tasks is provided in Appendix~\ref{sec:select_task}.

\paragraph{Baselines.}
We compare \toolName{} with production coding agent harnesses, both in their default mode and with their native subagents enabled, as well as with a research centralized multi-agent system.
\textbf{Codex} (v0.153.4)~\citep{openai_codex_2025} and \textbf{Claude Code} (v2.1.280)~\citep{anthropic_claude_code_2026} are agentic coding harnesses that enable language models to work on coding tasks using tools for file editing and shell execution.
\textbf{Codex (subagent)} and \textbf{Claude Code (subagent)} use the same harnesses, with one addition: the task prompt ends with the instruction ``Use [Codex/Claude Code]'s native subagents to help solve this task.'' This lets the main agent spawn and coordinate sub-agents through each harness's built-in mechanism.
\textbf{AOrchestra}~\citep{ruan2026aorchestra} is a centralized multi-agent system. Its main orchestrator creates specialized sub-agents on demand and instantiates each one as a tuple of instruction, context, tools, and model.
On SWE-bench Verified, we also compare with \textbf{mini-SWE-agent}~\citep{yang2024sweagent}, a lightweight software-engineering agent that uses a simple linear loop with Bash as its main interface.

\paragraph{Models and implementation.}
On Terminal-Bench 4.0 and DeepSWE v1.1, we use two frontier base models: GPT-6-Astra~\citep{openai2026gpt6astra} and Claude Opus 5.5~\citep{anthropic2026opus55}, both with \texttt{xhigh} reasoning effort. We build \toolName{} on top of Codex for GPT-6-Astra and on top of Claude Code for Claude Opus 5.5. Every baseline uses the same base model as \toolName{} in its setting, so the only difference between them is how agents coordinate. On SWE-bench Verified, all methods use Gemini 3 Flash~\citep{doshi2025gemini3flash}, accessed through OpenRouter with default reasoning and sampling parameters. For this setting, we route the Claude Code CLI through OpenRouter to support the non-Anthropic model. For every task, we evaluate \toolName{} with $n \in \{2, 4\}$ parallel agents.

\paragraph{Metrics.}
We report task accuracy, wall-clock latency, and cost. Wall-clock latency is the time from the first API call to the last generated token. We use two cost measures: Cost/Task is the total API cost of all agents in a single run, and Cost/Submission divides this total by the number of agents, since each \toolName{} agent produces its own submission that is scored independently. All results are averaged over three runs and reported as mean $\pm$ standard deviation.

\subsection{Terminal-Bench 4.0}
\label{sec:results-tb40}

We evaluate on the 10 long-horizon Terminal-Bench 4.0 tasks described in Appendix~\ref{sec:select_task}. Table~\ref{tab:terminalbench} shows that \toolName{} improves both accuracy and latency under both base models. With GPT-6-Astra, \toolName{} ($n=2$) reaches 85.00\% accuracy, 13.3 points higher than the strongest baseline (Codex and Codex with subagents, both 71.67\%), and runs $1.54\times$ faster than Codex, at a higher Cost/Task (\$13.65 vs.\ \$9.01) but a lower Cost/Submission (\$6.83). With Claude Opus 5.5, \toolName{} ($n=2$) reaches 93.33\%, 11.7 points above Claude Code, while cutting latency from 130.75 to 64.77 minutes ($2.02\times$) at a lower Cost/Task (\$17.91 vs.\ \$20.31).

Increasing to $n{=}4$ gives the largest speedups ($2.05\times$ and $2.49\times$): more work proceeds in parallel, and findings reach peers earlier. It also gives the lowest Cost/Submission, since each agent does a smaller share of the work and total cost grows more slowly than the number of submissions. Accuracy drops by 3--4 points relative to $n{=}2$, because each agent still submits a complete solution while building less of it itself. More of each solution is assembled from peers' components, which raises the risk of integration errors and lets a defect in one shared result affect several submissions. $n{=}4$ nonetheless remains above every baseline.

Native subagent baselines give model-dependent trade-offs in accuracy and latency, but they consistently raise cost. Codex (subagent) matches vanilla Codex in accuracy and runs 1.34$\times$ faster, but at 2.9$\times$ the cost (\$25.80 vs.\ \$9.01). Claude Code with subagents lowers accuracy from 81.67\% to 80.00\% while increasing latency and raising cost by 1.5$\times$ (\$30.88 vs.\ \$20.31). The trajectories show that delegation adds work rather than dividing it. Subagents unaware of one another repeat parts of the same investigation; and the main agent adds further rounds of testing and review while remaining responsible for integration and final delivery. 
In contrast, \toolName{} ($n{=}2$) costs less per task than either subagent baseline (\$13.65 and \$17.91) because it keeps communication cheap and makes knowledge reuse direct.

\begin{table}[H]
\centering
\small
\setlength{\tabcolsep}{5pt}
\renewcommand{\arraystretch}{1.15}
\caption{Comparison on Terminal-Bench 4.0 across two base models.
Best mean values within each model are bolded. Cost/Task sums the costs of all agents in a single run. Cost/Submission divides this total by the number of agents, as each agent produces one submission that is scored independently.}
\label{tab:terminalbench}
\begin{threeparttable}
\scalebox{1}{
\centering
\begin{tabular}{@{}l c c | c c@{}}
\toprule
\textbf{Method}
& \textbf{Avg. Acc (\%)}
& \textbf{Avg. Latency}
& \textbf{Cost/Task}
& \textbf{Cost/Submission} \\
\midrule
\multicolumn{5}{@{}l}{\textit{GPT-6-Astra}}\\
Codex
& $71.67 \pm 11.55$
& 26.91 $\pm$ 0.91 min ($1.00\times$)
& \textbf{\$9.01} $\pm$ \$0.71
& \$9.01 $\pm$ \$0.71 \\
Codex (subagent)
& $71.67 \pm 11.69$
& 20.04 $\pm$ 1.94 min ($1.34\times$)
& \$25.80 $\pm$ \$2.21
& \$25.80 $\pm$ \$2.21 \\
AOrchestra & $70.00 \pm 5.00$
&  25.78 $\pm$ 1.15 min ($1.04\times$)
& \$11.94 $\pm$ \$1.02
& \$11.94 $\pm$ \$1.02 \\
\rowcolor{rowgray}\textbf{\toolName{} ($n=2$)}
& $\mathbf{85.00} \pm 13.23$
& 17.47 $\pm$ 1.82 min ($1.54\times$)
& {\$13.65} $\pm$ \$1.49
& \$6.83 $\pm$ \$0.74 \\
\rowcolor{rowgray}\textbf{\toolName{} ($n=4$)}
& $81.67 \pm 12.33$
& \textbf{13.16} $\pm$ 1.43 min ($\mathbf{2.05\times}$)
& \$23.90 $\pm$ \$2.27
& \textbf{\$5.97} $\pm$ \$0.57 \\
\midrule
\multicolumn{5}{@{}l}{\textit{Claude Opus 5.5}}\\
Claude Code
& $81.67 \pm 2.89$
& 130.75 $\pm$ 6.27 min ($1.00\times$)
& \$20.31 $\pm$ \$0.23
& \$20.31 $\pm$ \$0.23 \\
Claude Code (subagent)
& $80.00 \pm 5.00$ & $138.58 \pm 10.17$ min ($0.94\times$) & $\$30.88 \pm \$3.17$ & $\$30.88 \pm \$3.17$  \\
AOrchestra & $76.67 \pm 5.77$
& 157.21 $\pm$ 9.03 min ($0.83\times$)
& \$24.22 $\pm$ \$1.05
& \$24.22 $\pm$ \$1.05\\
\rowcolor{rowgray}\textbf{\toolName{} ($n=2$)}
& $\mathbf{93.33} \pm \text{3.33}$
& 64.77 $\pm$ 5.03 min ($2.02\times$)
& \textbf{\$17.91} $\pm$ \$1.17
& \$8.96 $\pm$ \$0.59 \\
\rowcolor{rowgray}\textbf{\toolName{} ($n=4$)}
& $89.17 \pm 5.20$
& \textbf{52.51} $\pm$ 4.43 min ($\mathbf{2.49\times}$)
& \$29.98 $\pm$ \$0.91
& \textbf{\$7.50} $\pm$ \$0.23 \\
\bottomrule
\end{tabular}}
\end{threeparttable}
\end{table}

\subsection{DeepSWE v1.1}
\label{sec:results-deepswe11}

We evaluate on the 10 long-horizon DeepSWE v1.1 tasks listed in Appendix~\ref{sec:select_task}. Table~\ref{tab:deepswe} shows that \toolName{} again achieves the best accuracy and latency under both base models. With GPT-6-Astra, \toolName{} ($n{=}2$) reaches 98.33\% accuracy, 8.3 points above the strongest baseline, Codex with subagents (90.00\%), and runs 1.25$\times$ faster than Codex. It also costs less per task than Codex with subagents (\$16.83 vs.\ \$24.97). With Claude Opus 5.5, the gains are larger: \toolName{} ($n{=}4$) reaches 90.83\%, 17.5 points above the strongest baseline, AOrchestra (73.33\%), and runs 1.57$\times$ faster than Claude Code. \toolName{} ($n{=}2$) reaches 88.33\% at roughly the same Cost/Task as Claude Code (\$12.15 vs. \$12.01).

Native subagents hurt Claude Code most: accuracy falls from 71.67\% to 50.00\%, while latency and cost increase by 15.9\% and 65.1\%, respectively. 
Trajectory analysis shows that most of its failed runs ended with an empty patch or a benchmark timeout, often because the main agent held the final commit while waiting on subagents that were cut off by the CLI's idle limit\footnote{Claude Code defaults to a 600-second limit on continuous idle waiting for background subagents; once it is reached, unfinished subagents are terminated and the process exits~\citep{anthropic_claudecode_headless}.} or that prolonged integration with extra review rounds.
By comparison, \toolName{} (n=2) achieves higher accuracy and lower latency and cost than the corresponding native subagent baselines. Its shared context makes recorded findings and failed attempts directly accessible to peer agents. These results highlight that the value of parallel work depends on how effectively it is reused and converted into completed submissions, rather than on delegation alone.
\begin{table}[H] 
\centering
\small
\setlength{\tabcolsep}{5pt}
\renewcommand{\arraystretch}{1.15}
\caption{Comparison on DeepSWE v1.1 across two base models.
Best mean values within each model are bolded. Cost/Task sums the costs of all agents in a single run. Cost/Submission divides this total by the number of agents, as each agent produces one submission that is scored independently.}
\label{tab:deepswe}
\begin{threeparttable}[H]
\scalebox{1}{
\centering
\begin{tabular}{@{}l c c | c c@{}}
\toprule
\textbf{Method}
& \textbf{Avg. Acc (\%)}
& \textbf{Avg. Latency}
& \textbf{Cost/Task}
& \textbf{Cost/Submission} \\
\midrule
\multicolumn{5}{@{}l}{\textit{GPT-6-Astra}}\\
Codex
& $88.33 \pm 2.89$
& 16.57 $\pm$ 0.39 min ($1.00\times$)
& \textbf{\$9.40} $\pm$ \$0.18
& \$9.40 $\pm$ \$0.18 \\
Codex (subagent)
& $90.00 \pm 0.00$
& 15.76 $\pm$ 2.05 min ($1.05\times$)
& \$24.97 $\pm$ \$3.15
& \$24.97 $\pm$ \$3.15 \\
AOrchestra & $83.33 \pm 7.64$
& 18.31 $\pm$ 1.08 min ($0.91\times$)
& {\$10.74} $\pm$ \$0.97
& {\$10.74} $\pm$ \$0.97 \\
\rowcolor{rowgray}\textbf{\toolName{} ($n=2$)}
& $\mathbf{98.33} \pm 2.89$
& 13.29 $\pm$ 0.98 min ($1.25\times$)
& {\$16.83} $\pm$ \$1.65
& \$8.42 $\pm$ \$0.82 \\
\rowcolor{rowgray}\textbf{\toolName{} ($n=4$)}
& $90.00 \pm 10.00$
& \textbf{12.37} $\pm$ 0.66 min ($\mathbf{1.34\times}$)
& \$30.87 $\pm$ \$1.09
& \textbf{\$7.72} $\pm$ \$0.27 \\
\midrule
\multicolumn{5}{@{}l}{\textit{Claude Opus 5.5}}\\
Claude Code
& $71.67 \pm 2.89$
& 49.58 $\pm$ 3.11 min ($1.00\times$)
& \textbf{\$12.01} $\pm$ \$0.60
& \$12.01 $\pm$ \$0.60 \\
Claude Code (subagent)
& $50.00 \pm 10.00$ & $57.46 \pm 2.16$ min ($0.86\times$) & $\$19.83 \pm \$0.99$ & $\$19.83 \pm \$0.99$ \\
AOrchestra & $73.33 \pm 5.77$
& 47.32 $\pm$ 2.86 min ($1.05\times$)
& \$12.11 $\pm$ \$0.52
& \$12.11 $\pm$ \$0.52 \\
\rowcolor{rowgray}\textbf{\toolName{} ($n=2$)}
& $88.33 \pm 2.89$
& 32.98 $\pm$ 2.47 min ($1.50\times$)
& {\$12.15} $\pm$ \$0.57
& \textbf{\$6.07} $\pm$ \$0.28 \\
\rowcolor{rowgray}\textbf{\toolName{} ($n=4$)}
& $\mathbf{90.83} \pm 8.78$
& \textbf{31.61} $\pm$ 0.96 min ($\mathbf{1.57\times}$)
& \$24.49 $\pm$ \$0.67
& \$6.12 $\pm$ \$0.17 \\
\bottomrule
\end{tabular}}
\end{threeparttable}
\end{table}

\paragraph{Coordination with a low-cost model.}
We additionally evaluate DeepSeek-V4.1-Flash~\citep{deepseekai2026deepseekv41flashpushinglimitskv} with Codex on the
same 10 long-horizon DeepSWE tasks.
Table~\ref{tab:deepswe-deepseek} shows that \toolName{} ($n=2$)
improves mean accuracy from 76.67\% to 88.33\% (+11.7 points)
and reduces mean latency from 27.67 to 22.25 minutes (1.24$\times$ faster).
Notably, \toolName{} with DeepSeek-V4.1-Flash matches the mean accuracy of vanilla Codex with GPT-6-Astra (88.33\%) at approximately $1/28$ of its total cost per task (\$0.33 vs.\ \$9.40), though it remains slower (22.25 vs.\ 16.57 minutes), consistent with the
more frequent revisions and failed attempts observed in its trajectories.
These results suggest that decentralized coordination can
make a low-cost model more competitive on long-horizon tasks.
\begin{table}[t]
\centering
\caption{Comparison of Codex and \toolName{} with DeepSeek-V4.1-Flash on the same 10 long-horizon DeepSWE v1.1 tasks.
Results are mean $\pm$ standard deviation across three repeats.}
\label{tab:deepswe-deepseek}
\small
\setlength{\tabcolsep}{6pt}
\begin{tabular}{lcccc}
\toprule
\textbf{Method}
& \textbf{Avg. Acc (\%)}
& \textbf{Avg. Latency}
& \textbf{Cost/Task}
& \textbf{Cost/Submission} \\
\midrule
{\textit{DeepSeek-V4.1-Flash}}\\
Codex
& $76.67 \pm 11.55$
& $27.67 \pm 2.17$ min ($1.00\times$)
& $\$\mathbf{0.19} \pm \$0.01$
& $\$0.19 \pm \$0.01$ \\
\rowcolor{gray!12}
\textbf{\toolName{} ($n=2$)}
& $\mathbf{88.33} \pm 7.64$
& $\mathbf{22.25} \pm 0.90$ min ($\mathbf{1.24}\times$)
& $\$0.33 \pm \$0.01$
& $\$\mathbf{0.17} \pm \$0.01$ \\
\bottomrule
\end{tabular}
\end{table}

\subsection{SWE-bench Verified}
\label{sec:swe}

SWE-bench Verified tests whether \toolName{} still helps on a widely used software-engineering benchmark with a lightweight base model. We build \toolName{} based on AOrchestra's harness and compare it with Claude Code, mini-SWE-agent, and AOrchestra, all using Gemini 3 Flash.
Table~\ref{tab:swebench_verified} shows that \toolName{} outperforms all baselines in both accuracy and latency. With $n=4$, \toolName{} reaches 66.08\% accuracy, 10.8 points higher than the strongest baseline, AOrchestra (55.26\%). It is also $2.12\times$ faster than Claude Code and $1.47\times$ faster than AOrchestra. With $n=2$, \toolName{} still improves accuracy by 8.5 points over AOrchestra at the same Cost/Task (\$0.24), which halves the cost per submission. Going from $n=2$ to $n=4$ roughly doubles Cost/Task but leaves Cost/Submission unchanged at \$0.12, and it further improves both accuracy and latency. 

These results show that multiple agents help even on largely sequential tasks. A single agent rarely completes every step correctly on its first attempt. It often inspects irrelevant files, pursues a wrong root cause, or submits a patch that fails the tests before it recovers. Each such detour costs time, and a single agent can only discover it by paying that cost itself. In \toolName{}, agents explore different paths in parallel and record what they learn in the shared context. Once one agent has hit a dead end, the others can skip it, and once one has localized a fault, the others can build on that finding. This shortens the path to a correct fix, which explains why \toolName{} is both faster and more accurate. A more detailed analysis is provided in Appendix~\ref{sec:swe_analyze}.

\begin{table}[t]
\centering
\small
\setlength{\tabcolsep}{5pt}
\renewcommand{\arraystretch}{1.15}
\caption{Comparison on SWE-bench Verified with Gemini 3 Flash. Best mean values within each model are bolded. Cost/Task sums the costs of all agents in a single run. Cost/Submission divides this total by the number of agents, as each agent produces one submission that is scored independently.}
\label{tab:swebench_verified}
\begin{threeparttable} [H]
\begin{tabular}{@{}l c c | c c@{}}
\toprule
\textbf{Method}
& \textbf{Avg. Acc (\%)}
& \textbf{Avg. Latency}
& \textbf{Cost/Task}
& \textbf{Cost/Submission} \\
\midrule
\multicolumn{5}{@{}l}{\textit{Gemini 3 Flash}}\\
Claude Code
& $49.32 \pm 1.98$
& 129.13 $\pm$ 5.12 s ($1.00\times$)
& {\$1.00} $\pm$ \$0.00\,\tnote{a}
& \$1.00 $\pm$ \$0.00\,\tnote{a} \\
mini-SWE-agent
& $54.73 \pm 2.37$
& 91.78 $\pm$ 3.53 s ($1.41\times$)
& {\$0.26} $\pm$ \$0.07
& \$0.26 $\pm$ \$0.07 \\
AOrchestra & $55.26 \pm 2.03$
& 89.94 $\pm$ 3.07 s ($1.44\times$)
& \textbf{\$0.24} $\pm$ \$0.04
& {\$0.24} $\pm$ \$0.04 \\
\rowcolor{rowgray}\textbf{\toolName{} ($n=2$)}
& ${63.72} \pm 2.29$
& 71.86 $\pm$ 2.88 s ($1.80\times$)
& \textbf{\$0.24} $\pm$ \$0.05
& \textbf{\$0.12}  $\pm$ \$0.03 \\
\rowcolor{rowgray}\textbf{\toolName{} ($n=4$)}
& $\mathbf{66.08} \pm 1.86$
& \textbf{61.01} $\pm$ 2.63 s ($\mathbf{2.12\times}$)
& \$0.47 $\pm$ \$0.07
& \textbf{\$0.12} $\pm$ \$0.02 \\
\bottomrule
\end{tabular}
\begin{tablenotes}[flush]
\footnotesize
\item[a] The Claude Code CLI sends \texttt{cache\_control} blocks in the Anthropic API format, so cache reuse only works if the upstream provider honors those blocks. For Gemini-3-Flash the real cost without cache reuse is therefore around \$1 per task.
\end{tablenotes}
\end{threeparttable}
\end{table}

\subsection{ProgramBench}\label{sec:program-bench}
\begin{figure}[t]
\centering
\includegraphics[width=1.0\linewidth]{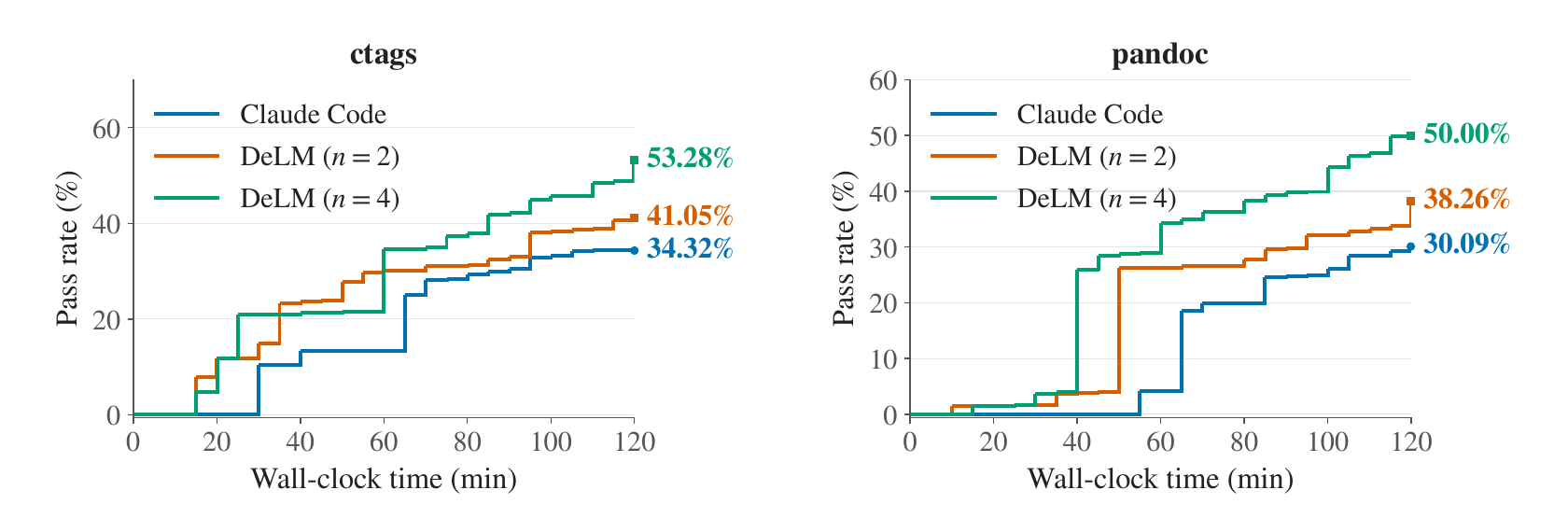}
\caption{\textbf{\toolName{} makes more progress within the same time budget.}
Test pass rate over time on two ProgramBench tasks with Claude Opus 5.5
under a 120-minute budget. \toolName{} uses Claude Code as its per-agent harness. From 60 minutes onward, the curves stay ordered ($n{=}4$ above $n{=}2$ above Claude Code), so \toolName{} reaches any given pass rate sooner and ends the budget higher.}
\label{fig:programbench}
\end{figure}

ProgramBench asks agents to rebuild an entire program from its compiled
executable and documentation. There is no natural stopping point: the
pass rate keeps rising for as long as agents keep implementing. We
therefore use it to ask a different question from the preceding
sections: given a fixed wall-clock budget, how much progress does each
system make, and how quickly? We run Claude Code and \toolName{}, built on
top of Claude Code, with Claude Opus 5.5 (\texttt{xhigh} reasoning
effort) on two hard tasks, \texttt{ctags} and \texttt{pandoc}. Each run
has a 120-minute budget, and we track the hidden-test pass rate over
time (Figure~\ref{fig:programbench}).

\paragraph{Higher pass rates under the same budget.}
At the end of the budget, \toolName{} ($n{=}2$) improves over Claude Code
by 6.73 points on \texttt{ctags} (41.05\% vs.\ 34.32\%) and by 8.17
points on \texttt{pandoc} (38.26\% vs.\ 30.09\%). With $n{=}4$, the
gains grow to 18.96 and 19.91 points (53.28\% and 50.00\%). Each
\toolName{} agent runs the same Claude Code harness and model as the
baseline. The gains therefore come from the coordination layer, not
from a stronger individual agent.

\paragraph{Faster progress toward the same result.}
The advantage is not limited to the final checkpoint. From 60 minutes
onward, the curves stay ordered on both tasks, with $n{=}4$ above
$n{=}2$ above Claude Code, so \toolName{} reaches any given pass rate
sooner. On \texttt{ctags}, Claude Code reaches its final 34.32\% only
at the end of the budget. \toolName{} matches it after about 95 minutes
with $n{=}2$ and after about 60 minutes with $n{=}4$, which is half the
budget. Decentralized coordination thus turns each additional agent
into faster progress, not only a better final result.

\subsection{Ablation Study}\label{sec:ablation}
\begin{wrapfigure}{r}{0.45\linewidth}
    \centering
    \includegraphics[width=0.9\linewidth]{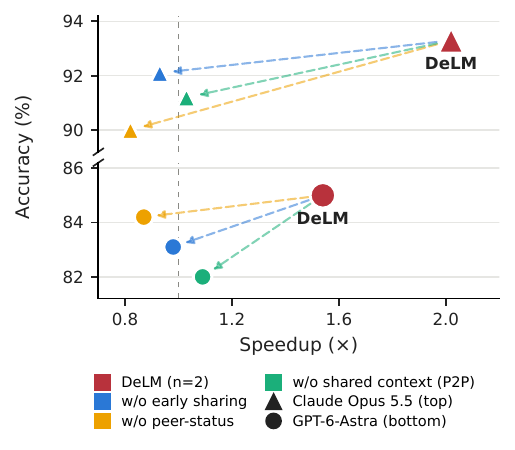}
    \caption{\textbf{Ablation on Terminal-Bench 4.0.} Arrows point from \toolName{} ($n=2$) to each variant with one mechanism removed. }
    \label{fig:ablation}
\end{wrapfigure}

To isolate the contribution of each coordination mechanism in Table~\ref{tab:agent-comparison}, we remove one mechanism at a time from \toolName{} ($n=2$) on Terminal-Bench 4.0 and keep everything else fixed. \textbf{w/o early sharing} lets agents publish results only after they complete their claimed subtask. \textbf{w/o peer-status} hides item owners and session status, so agents cannot see what their peers are currently working on. \textbf{w/o shared context} removes the shared context entirely, and agents coordinate by sending messages directly to each other. Figure~\ref{fig:ablation} shows the results.

The most striking effect is on latency. With the full system, the two agents run $1.54\times$ (GPT-6-Astra) and $2.02\times$ (Claude Opus 5.5) faster than the single-agent harness, whereas every ablation falls to $0.82$--$1.09\times$. Two parallel agents are thus no faster than one unless they coordinate effectively, so the speedup of \toolName{} comes from coordination rather than from parallelism alone. 

Removing peer-status awareness causes the largest slowdown under both models ($0.87\times$ and $0.82\times$), and with Claude Opus 5.5 it also causes the largest accuracy drop ($-3.3$ points). Without knowing what their peers are working on, agents tend to claim overlapping work and discover the duplication only after both have spent time on it. Removing early sharing brings the speedup back to about $1\times$ ($0.98\times$ and $0.93\times$), because an agent whose work depends on a peer's output must wait until the peer finishes its entire subtask instead of starting from partial results. Replacing the shared context with direct messaging retains slightly more speed than the other ablations ($1.09\times$ and $1.03\times$) but causes the largest accuracy drop with GPT-6-Astra ($-3.0$ points). Direct messages still let agents exchange information, but they do not accumulate into a persistent record that agents can revisit, build on, or correct, so findings and fixes are more easily missed.
\section{Trace Analysis: How Does Decentralized Coordination Help?}
\label{sec:trace}

To understand how the coordination mechanisms in Table~\ref{tab:agent-comparison} squeeze out bubbles, we analyze all \toolName{} runs on Terminal-Bench 4.0 and DeepSWE v1.1. Parallel subtask execution lets agents build components concurrently (\S\ref{sec:trace-analysis-parallel}). Intermediate-results sharing lets agents start dependent work before a peer finishes, avoiding relay waits (\S\ref{sec:trace-analysis-early-sharing}). Peer-status awareness keeps agents from duplicating a peer's search, avoiding redundant work (\S\ref{sec:trace-analysis-peer-awareness}). Finally, a stateful shared context lets corrections reach every agent without another round of delegation, while staying small enough to read cheaply (\S\ref{sec:trace-analysis-reuse}).

\subsection{Parallel subtask execution shortens the critical path}
\label{sec:trace-analysis-parallel}


Parallel agents speed up a task only if the work is divided so that no agent idles or duplicates a peer. \toolName{} agents claim distinct components from the task queue, share their code through the shared context, and each integrates its peers' components into its own solution. Adding agents makes this division finer and shortens the critical path.
\begin{figure}[t]
\centering
\includegraphics[width=0.9\linewidth]{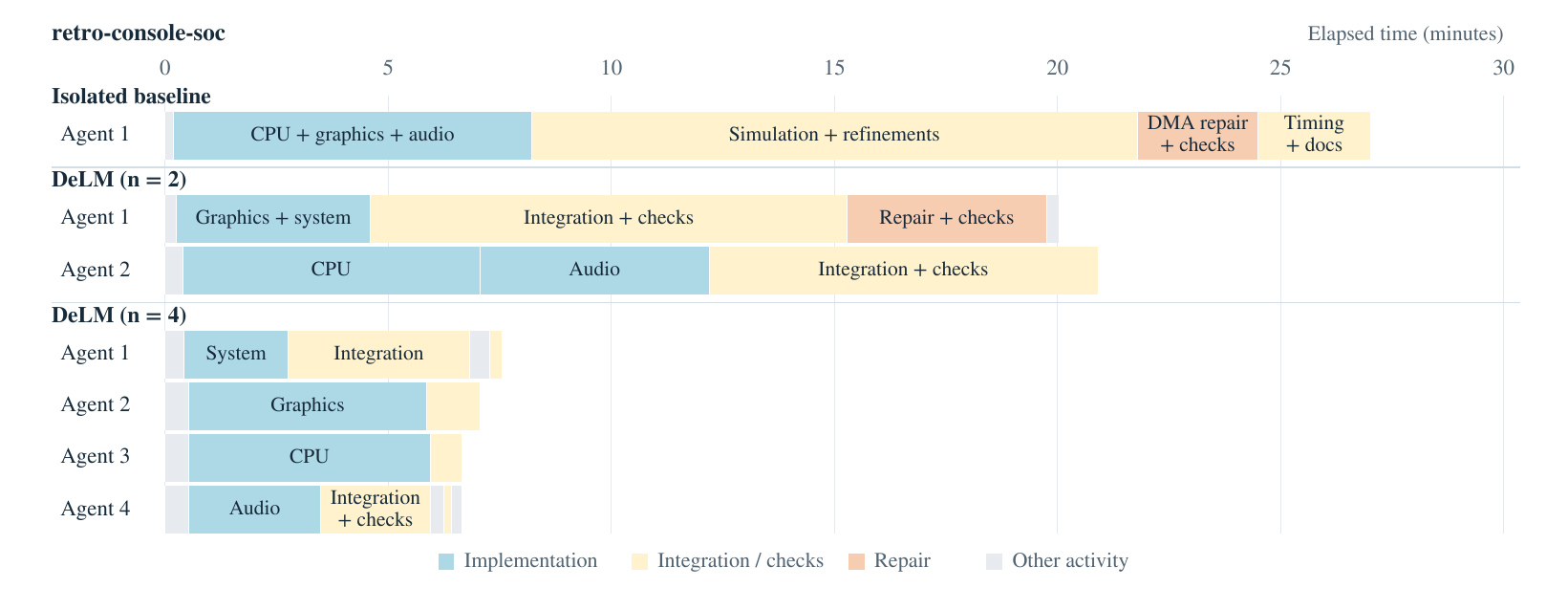}
\caption{\textbf{More agents build more components in parallel.} Execution timelines on Terminal-Bench 4.0's \texttt{retro-console-soc} task with GPT-6-Astra.}
\label{fig:retro-console-soc-timeline}
\end{figure}

Figure~\ref{fig:retro-console-soc-timeline} shows this on Terminal-Bench’s \texttt{retro-console-soc} task with GPT-6-Astra. A single agent builds the
CPU, graphics, and audio in sequence and finishes in 27.0 minutes. With
$n{=}2$, one agent builds the graphics and system components while its
peer builds the CPU and then the audio, finishing in 20.9 minutes
($1.29\times$). With $n{=}4$, separate agents build the system,
graphics, CPU, and audio concurrently, and the task finishes in 7.5
minutes ($3.6\times$). The same pattern appears across Terminal-Bench
4.0: moving from $n{=}2$ to $n{=}4$ raises the average speedup from
$1.54\times$ to $2.05\times$ with GPT-6-Astra and from $2.02\times$ to
$2.49\times$ with Claude Opus 5.5 (\S\ref{sec:results-tb40}).

\subsection{Intermediate-results sharing unblocks dependent work}
\label{sec:trace-analysis-early-sharing}

When one component depends on another, centralized orchestration makes
the consumer wait until the producer finishes its subtask and the main
agent relays the result. \toolName{} agents instead publish partial results as soon as they are usable, so dependent work and integration testing overlap with implementation rather than following it.

In Terminal-Bench's \texttt{telecom-entity-resolution} task, Agent 2's household-separation step consumes the clusters produced by Agent 1's record matcher. Agent 1 shared the matcher as soon as it ran, before its thresholds were tuned, so Agent 2 developed its step on real clusters while the matcher was still being refined.




\subsection{Peer-status awareness avoids duplicate exploration}
\label{sec:trace-analysis-peer-awareness}

Parallel agents that cannot see each other's work tend to converge on the same promising direction. In centralized systems, each sub-agent sees only the subcontext the main agent assigned it, so overlapping work goes unnoticed until the main agent collects the results. \toolName{} exposes every agent's claims, status, and stated scope through the task queue and shared context, so an agent can notice an overlap while its search is still running and redirect.

In Terminal-Bench's \texttt{wdm-design} task, agents optimize an optical device by changing its port geometry and material pattern. In one four-agent run, Agents 1 and 3 independently began tuning the same port widths and positions. Agent 1 then saw Agent 3's claim through the shared context:

\begin{lstlisting}[style=gist]
[agent-3/FACT] [...] I will tune only w_in,w_long,w_short,y_long,y_short [...]
[agent-1/FACT] [...] applying the cavity artifact revealed agent-3 had just claimed the same five-parameter port tuning. [...]
[agent-1/EXEC] kill -TERM 770
[agent-1/EXEC] python binary_refine.py [...]
\end{lstlisting}

Agent 1 terminates its own port search and returns to refining the material pattern of its earlier candidate, while Agent 3 continues port tuning. The agents resolve the overlap themselves, before either search finishes, without a main agent having to detect the duplication and re-plan.

\subsection{Stateful shared context enables reuse and correction}
\label{sec:trace-analysis-reuse}

\begin{table}[t]
\centering
\small
\setlength{\tabcolsep}{6pt}
\renewcommand{\arraystretch}{1.15}
\caption{Final shared-context size (thousands of tokens) and the cost of reading it as a percentage of total task cost.}
\label{tab:shared-context-cost}
\begin{tabular}{@{}l c cc c cc@{}}
\toprule
& & \multicolumn{2}{c}{\textbf{Terminal-Bench 4.0}} & & \multicolumn{2}{c}{\textbf{DeepSWE v1.1}} \\
\cmidrule{3-4} \cmidrule{6-7}
\textbf{Model} & \textbf{Agents}
& \textbf{Context (k)} & \textbf{Cost (\%)} &
& \textbf{Context (k)} & \textbf{Cost (\%)} \\
\midrule
\multirow{2}{*}{GPT-6-Astra}
& $n=2$ & $7.31 \pm 0.42$  & $9.17 \pm 0.38$  & & $7.54 \pm 0.21$  & $7.71 \pm 0.18$ \\
& $n=4$ & $12.88 \pm 0.97$ & $17.68 \pm 0.25$ & & $13.03 \pm 0.96$ & $14.56 \pm 0.37$ \\
\midrule
\multirow{2}{*}{Claude Opus 5.5}
& $n=2$ & $6.62 \pm 0.34$  & $1.92 \pm 0.11$  & & $4.67 \pm 0.11$  & $1.69 \pm 0.06$ \\
& $n=4$ & $14.48 \pm 0.47$ & $3.66 \pm 0.29$  & & $10.43 \pm 0.53$ & $3.04 \pm 0.23$ \\
\bottomrule
\end{tabular}
\end{table}


Shared results help only if later improvements reach the agents that already used them. Because updates and corrections are appended as
new entries that reference revised files, rather than overwriting old
ones (\S\ref{sec:shared-context}), an agent can re-import a peer's updated work at any time. Agents rely on this heavily. Across 720 agent trajectories on Terminal-Bench 4.0 and DeepSWE v1.1, agents imported shared files 5,237 times, and 80.3\% of trajectories later imported a revised version of a file they had already imported. Agents thus track their peers' work as it improves, and fixes reach every
agent that depends on them without re-planning. In centralized systems,
by contrast, correcting a result that has already been summarized and
relayed requires another round of delegation.

This reuse is inexpensive. Because agents publish short summaries and attach files by reference (§\ref{sec:shared-context}), the shared context averages under 15K tokens at task completion (Table~\ref{tab:shared-context-cost}). Each read also returns only what is new since the agent's last read, so most of this content enters an agent's session once. Accounting for all reads by all agents, including this cached reuse, reading the shared context accounts for 1.7--3.7\% of total task cost with Claude Opus 5.5 and 7.7--17.7\% with GPT-6-Astra.

\section{Related Work}
\label{sec:related_work}

\paragraph{Multi-agent coordination.}
Recent LLM-based multi-agent systems improve performance by decomposing tasks across specialized or interacting agents. CAMEL~\citep{li2023camel} instantiates role-playing agents to study autonomous cooperation, while MetaGPT~\citep{hong2023metagpt} and ChatDev~\citep{qian2024chatdev} organize collaboration through software-engineering roles and standardized workflows. More recent systems make orchestration adaptive. AOrchestra~\citep{ruan2026aorchestra} creates task-specific sub-agents on demand, ToolOrchestra~\citep{su2025toolorchestra} trains a lightweight orchestrator to coordinate models and tools, Squeeze Evolve~\citep{maheswaran2026squeeze} routes the stages of evolutionary inference to models of different cost and capability, and RecursiveMAS~\citep{yang2026recursive} replaces text-level communication with recursive latent-state transfer. These systems coordinate through either a fixed workflow or a central orchestrator that decides what each agent sees and when progress is merged. Combee~\citep{li2026combee} likewise identifies centralized aggregation as a bottleneck under high parallelism, but addresses it for parallel prompt learning rather than agent coordination. 

\paragraph{Blackboard and shared-memory architectures.} Blackboard architectures coordinate problem solvers through a shared workspace~\citep{erman1980hearsay, nii1986part}. Recent LLM variants use a blackboard on which agents post messages for general reasoning~\citep{han2025exploring} or to which agents volunteer information in response to requests for data discovery~\citep{salemi2025llm}. However, these systems still rely on a control agent to select which agent acts or which request is posted, and they treat the board mainly as a message channel. \toolName{} removes the controller from the loop entirely. Agents claim work asynchronously from a task queue, see what their peers are working on, and build on or correct earlier findings by appending to a stateful shared context (Table~\ref{tab:agent-comparison}). We further target long-horizon agentic coding, where shared progress includes repository and environment state rather than only textual messages.

\paragraph{Agentic coding harnesses.}
Language models become far more capable when coupled with executable environments that let them inspect state, invoke tools, and act on feedback. SWE-agent~\citep{yang2024sweagent} and its lightweight variant mini-SWE-agent show that a well-designed agent-computer interface substantially improves software-engineering performance. Production harnesses such as Claude Code~\citep{anthropic_claude_code_2026} and Codex~\citep{openai_codex_2025} let models read repositories, edit files, execute commands, and increasingly delegate work to native subagents. Recursive Language Models~\citep{zhang2025recursive} apply a similar code-mediated loop to long-context reasoning by treating the prompt as an external environment. \toolName{} complements these harnesses: rather than replacing the single-agent loop, it adds a coordination layer on top of it, and we instantiate it on both Codex and Claude Code.

\section{Discussion and Limitations}

\paragraph{Conclusion.} We introduced \toolName{}, a decentralized multi-agent framework in which agents coordinate through a shared context and a task queue rather than through a main agent. By letting agents claim work asynchronously, publish findings as soon as they are usable, and see what every peer is doing, \toolName{} squeezes out the bubbles that existing designs create: redundant work among independent agents, barrier waits among peer-communicating agents, and relay waits under centralized orchestration. Across Terminal-Bench 4.0, DeepSWE v1.1, and SWE-bench Verified, \toolName{} is both more accurate and faster than single-agent harnesses, their native subagents, and AOrchestra, and on ProgramBench it makes more progress than Claude Code under the same time budget. Our ablation shows that these gains come from coordination rather than parallelism alone: removing any single mechanism brings two agents back to roughly single-agent speed. These results suggest that scaling multi-agent systems requires not only more agents, but a coordination substrate that keeps every agent
productive while keeping the cost of communication among them low.

\paragraph{Limitations.}  In certain settings, lower latency and Cost/Submission for \toolName{} come with higher Cost/Task. A natural-language shared context is also less suited to exact aggregation over structured data, where \toolName{} can be enhanced when combined with code-mediated execution (Appendix~\ref{sec:with_rlm}).

\section{Acknowledgements}

We thank Shayan Talaei, Jon Saad-Falcon, Jacky Kwok, Hermann Kumbong, Ishan Khare, Miria Feng, Swapnil Gandhi, Michael Y. Li, Chun Deng, Chong Zeng, Rui Li, Ke Li, Marquita
Ellis, Adrian Gamarra Lafuente, Charles Ding, Tarun Suresh, Ziyu Chen, Zhuohan
Gu, Yize Liu, and Ligeng Zhu. We would like to thank our collaborators at the Stanford Artificial
Intelligence Laboratory (SAIL) and Stanford HAI. We are also
grateful to Justin Oh for his generous financial support of this research.\\

We gratefully acknowledge support from federal sources: NSF under No. 24-554 (AIMing) and
DARPA under No. HR00112520038 (Fallingwater). We also gratefully acknowledge support from
Stanford HAI, IBM (Code Generation), Lightspeed, Google, and Google DeepMind.
\newpage

\bibliography{bibliography}
\bibliographystyle{plainnat}

\newpage
\appendix

\section{Method Details}

\begin{algorithm}
\caption{\textbf{\toolName{} pipeline.} Agents coordinate concurrent work through a shared context $\sctx$ and task queue $\tque$. They publish intermediate results for others to reuse, update the queue as needed, and return their solutions independently.}
\label{alg:dlm}
\begin{algorithmic}[1]
\setlength{\itemsep}{0.6pt}
\Require Task $D$; number of agents $n$
\Ensure Agent solutions $\{Y_i\}_{i=1}^{n}$
\smallskip

\State $\sctx \gets \emptyset;\quad \tque \gets \{T_i\}_{i=1}^{n}$ \stage{1}

\ForAll{agents $i=1,\ldots,n$ \textbf{in parallel}}
    \State $r_i \gets \emptyset$ \Comment{accumulated local work}
    \Repeat
        \State $r_i \gets \Call{ContinueWork}{D, r_i, \tque, \sctx}$ \stage{2}
        \If{useful intermediate results or corrections are available}
            \State $G_i \gets \Call{Summarize}{r_i}$ \stage{3}
            \State \Call{Publish}{$\sctx, G_i$} \Comment{references to shared files}
        \EndIf
        \State \Call{UpdateSubtasks}{$\tque, r_i, \sctx$} \Comment{complete or add work} \stage{4}
    \Until{\Call{ReadyToSubmit}{$D, r_i, \sctx$}}
    \State $Y_i \gets \Call{Finalize}{D, r_i, \sctx}$ \stage{5}
\EndFor

\State \Return $\{Y_i\}_{i=1}^{n}$
\end{algorithmic}
\end{algorithm}

\subsection{Selective Retrieval and Reuse}
\label{sec:unfold}
A compact entry $G_i$ in the shared context $\sctx$ can reference files captured at publication time. These files remain available under the publication's identifier, even as the contributing agent continues editing its own workspace. Agents access them through two operations, \texttt{expand} and \texttt{apply}, which separate inspection from incorporation into a local solution.

\paragraph{Expansion (\texttt{expand}).}
The \texttt{expand} operation returns a publication's attached files as tool output without changing the agent's workspace. By default, it shows differences against the original task files, with full contents for new files or files whose originals are unavailable. Using \texttt{---only}, the agent can select particular files and view them in full, allowing it to examine the implementation details or supporting evidence relevant to its current work.

\paragraph{Application (\texttt{apply}).}
The \texttt{apply} operation copies all or selected attachments into the receiving agent's workspace. It transfers complete files rather than merging patches, checking local contents against the task originals and published versions before writing. If a selected file contains local changes it cannot account for, the import is refused and the agent must reconcile those changes. Expansion is optional: an agent can inspect a contribution before incorporating it or apply selected files directly.

Expanded content enters the requesting agent's session, while applied files become part of its local workspace. Neither operation inserts file contents into the compact shared-context view, so agents can inspect and reuse detailed contributions without requiring every other agent to read them.

\section{Selected Tasks}\label{sec:select_task}
\toolName{} is designed for long-horizon tasks, where multiple agents have enough time to share progress and split up the remaining work. On short tasks, a single agent often finishes before any coordination can pay off. We therefore evaluate on a subset of long-running tasks from Terminal-Bench 4.0 and DeepSWE v1.1. We first ran the Codex baseline with GPT-6-Astra (\texttt{xhigh} reasoning effort) on each benchmark and recorded the wall-clock latency of every task. From Terminal-Bench 4.0, we selected 10 tasks whose latency fell between 15 minutes and 1 hour. The upper bound excludes tasks that run so long that repeated evaluation would be impractical. From DeepSWE v1.1, we selected 10 tasks with latency of at least 10 minutes. The selection criterion is based only on the baseline's latency. Table~\ref{tab:selected_tasks} lists the selected tasks together with their Codex latencies.
\begin{table}[h]
\centering
\scriptsize
\setlength{\tabcolsep}{3.5pt}
\renewcommand{\arraystretch}{1.08}
\caption{Selected tasks from Terminal-Bench 4.0 and DeepSWE v1.1.
Baseline latency is the average execution time of the Codex baseline using GPT-6-Astra at xhigh reasoning effort.}
\begin{minipage}[t]{0.48\textwidth}
\centering
\caption*{\textbf{Terminal-Bench 4.0}}
\begin{tabular}{@{}p{0.68\linewidth} r@{}}
\toprule
\textbf{Task} & \textbf{Codex Latency} \\
\midrule
ks-solver-cpp
& 21.00 min \\

telecom-entity-resolution
& 24.23 min \\

vf2-speedup-networkx
& 23.42 min \\

biped-contact-dynamics
& 21.56 min \\

cumulative-layout-shift
& 39.34 min \\

retro-console-soc
& 27.28 min \\

rs-archive-clone
& 21.92 min \\

wdm-design
& 39.27 min \\

lake-temp-glm
& 32.40 min \\

payments-pipeline-fix
& 18.67 min \\
\midrule
\textbf{Average}
& \textbf{26.91 min} \\
\bottomrule
\end{tabular}
\end{minipage}
\hfill
\begin{minipage}[t]{0.48\textwidth}
\centering
\caption*{\textbf{DeepSWE v1.1}}
\begin{tabular}{@{}p{0.68\linewidth} r@{}}
\toprule
\textbf{Task} & \textbf{Codex Latency} \\
\midrule
oxvg-structural-selector-preservation
& 19.71 min \\

wasmi-trap-coredumps
& 14.55 min \\

happy-dom-abort-pending-body-reads
& 12.52 min \\

scriggo-method-declarations
& 17.35 min \\

dynamodb-toolbox-lazy-recursive-schemas
& 22.19 min \\

dynamodb-toolbox-conditional-attribute
& 16.78 min \\

opa-template-string-reconstruction
& 13.99 min \\

boa-hierarchical-evaluation-cancellation
& 13.28 min \\

numba-stencil-boundary-modes
& 15.91 min \\

pebble-durability-wait-apis
& 19.38 min \\
\midrule
\textbf{Average}
& \textbf{16.57 min} \\
\bottomrule
\end{tabular}
\end{minipage}
\label{tab:selected_tasks}
\end{table}

\section{Experiments on LongBench-v2 Multi-Doc QA}
\label{sec:longbench}
\begin{table}[t]
\centering
\setlength{\tabcolsep}{5pt}
\renewcommand{\arraystretch}{1.2}
\caption{Accuracy (\%) comparison on LongBench-v2 Multi-Doc QA across four base models. Results are mean$\pm$std over 3 independent runs. Numbers in parentheses denote the number of samples in each domain. Best values are bolded.}
\label{tab:multidoc_results}
\begin{tabular}{@{}l ccccc | c@{}}
\toprule
\textbf{Method} & \textbf{Fin.} (15) & \textbf{Govern.} (23) & \textbf{MNews} (23) & \textbf{Legal} (14) & \textbf{Acad.} (50) & \textbf{Avg.} \\
\midrule
\multicolumn{7}{@{}l}{\textit{GPT-5.4}}\\
Base & 62.2\tiny{$\pm$9.3} & 39.1\tiny{$\pm$1.7} & 50.5\tiny{$\pm$8.6} & {62.0}\tiny{$\pm$4.4} & 56.0\tiny{$\pm$1.2} & 53.9\tiny{$\pm$4.9} \\
Claude Code & {65.0}\tiny{$\pm$8.3} & 36.2\tiny{$\pm$2.0} & 50.7\tiny{$\pm$2.0} & 59.5\tiny{$\pm$3.4} & \textbf{60.7}\tiny{$\pm$0.9} & 54.4\tiny{$\pm$3.1} \\
ReadAgent & 
{57.8}\tiny{$\pm$3.9} & 34.8\tiny{$\pm$5.0} & 52.2\tiny{$\pm$8.7} & 
{61.9}\tiny{$\pm$4.4} &
58.0\tiny{$\pm$1.2} &
53.0\tiny{$\pm$3.3} \\
\rowcolor{rowgray}\textbf{\toolName{}} & \textbf{71.1}\tiny{$\pm$3.1} & \textbf{43.5}\tiny{$\pm$2.0} & \textbf{65.2}\tiny{$\pm$2.0} & \textbf{64.3}\tiny{$\pm$5.8} & {56.0}\tiny{$\pm$2.8} & \textbf{60.1}\tiny{$\pm$1.2} \\
\midrule
\multicolumn{7}{@{}l}{\textit{Claude Sonnet 4.6}}\\
Base & 68.9\tiny{$\pm$3.9} & 34.8\tiny{$\pm$1.1} & 46.4\tiny{$\pm$2.0} & 52.4\tiny{$\pm$3.4} & 64.0\tiny{$\pm$0.9} & 53.3\tiny{$\pm$1.9} \\
Claude Code & 66.7\tiny{$\pm$1.1} & 42.0\tiny{$\pm$2.0} & 47.8\tiny{$\pm$4.4} & 50.0\tiny{$\pm$3.4} & 52.0\tiny{$\pm$3.4} & 51.7\tiny{$\pm$2.4} \\
ReadAgent & 62.2\tiny{$\pm$3.9} & {44.9}\tiny{$\pm$4.4} & 43.5\tiny{$\pm$4.4} & 57.1\tiny{$\pm$4.1} & \textbf{64.7}\tiny{$\pm$2.3} & 54.5\tiny{$\pm$3.8} \\
\rowcolor{rowgray}\textbf{\toolName{}} & \textbf{73.3}\tiny{$\pm$3.1} & \textbf{46.4}\tiny{$\pm$2.0} & \textbf{55.1}\tiny{$\pm$2.0} & \textbf{59.5}\tiny{$\pm$3.4} & \textbf{64.7}\tiny{$\pm$0.9} & \textbf{59.8}\tiny{$\pm$1.5} \\
\midrule
\multicolumn{7}{@{}l}{\textit{Gemini 3 Flash}}\\
Base & {80.0}\tiny{$\pm$0.0} & 30.4\tiny{$\pm$0.0} & 43.5\tiny{$\pm$0.0} & \textbf{71.4}\tiny{$\pm$0.0} & 60.0\tiny{$\pm$2.3} & 57.1\tiny{$\pm$0.5} \\
Claude Code & 73.3\tiny{$\pm$3.1} & 26.1\tiny{$\pm$1.5} & 39.1\tiny{$\pm$0.0} & 57.1\tiny{$\pm$2.1} & 52.0\tiny{$\pm$1.9} & 49.5\tiny{$\pm$2.2} \\
ReadAgent & 75.6\tiny{$\pm$2.3} & 21.7\tiny{$\pm$4.4} & 46.4\tiny{$\pm$2.9} & 69.0\tiny{$\pm$3.4} & 62.0\tiny{$\pm$2.3} & 54.9\tiny{$\pm$1.1} \\
\rowcolor{rowgray}\textbf{\toolName{}} & \textbf{82.2}\tiny{$\pm$3.1} & \textbf{37.8}\tiny{$\pm$1.6} & \textbf{52.2}\tiny{$\pm$1.5} & \textbf{71.4}\tiny{$\pm$0.0} & \textbf{64.0}\tiny{$\pm$0.0} & \textbf{61.5}\tiny{$\pm$0.6} \\
\midrule
\multicolumn{7}{@{}l}{\textit{DeepSeek-V4-Pro}}\\
Base & \textbf{86.7}\tiny{$\pm$3.1} & 47.8\tiny{$\pm$2.0} & 44.9\tiny{$\pm$2.0} & \textbf{76.2}\tiny{$\pm$3.4} & 64.0\tiny{$\pm$0.9} & 63.9\tiny{$\pm$1.3} \\
Claude Code & 84.4\tiny{$\pm$3.1} & 36.2\tiny{$\pm$2.0} & 56.5\tiny{$\pm$4.4} & 64.3\tiny{$\pm$2.1} & 48.7\tiny{$\pm$0.9} & 58.0\tiny{$\pm$0.9} \\
ReadAgent & 71.1\tiny{$\pm$3.9} & 43.5\tiny{$\pm$4.4} & 30.4\tiny{$\pm$2.3} & 38.1\tiny{$\pm$4.1} & 50.7\tiny{$\pm$1.2} & 46.8\tiny{$\pm$1.8} \\
\rowcolor{rowgray}\textbf{\toolName{}} & \textbf{86.7}\tiny{$\pm$5.4} & \textbf{52.2}\tiny{$\pm$2.0} & \textbf{59.4}\tiny{$\pm$4.4} & 73.8\tiny{$\pm$3.4} & \textbf{65.3}\tiny{$\pm$0.9} & \textbf{67.5}\tiny{$\pm$1.2} \\
\bottomrule
\end{tabular}
\end{table}

{LongBench-v2}~\citep{bai2025longbench} is a long-context benchmark designed to evaluate deep understanding and reasoning over realistic tasks. We focus on its multi-document QA setting, which contains 125 samples across diverse domains: Financial (15 samples), Government (23 samples), Multi-News (23 samples), Legal (14 samples), and Academic (50 samples). This setting is also suitable for our evaluation because its questions require multi-hop reasoning over evidence distributed across long contexts, rather than extracting a single local span.

For the compared baselines, we additionally pick {ReadAgent}~\citep{lee2024human} which is an agentic long-context baseline that compresses the input into short gist memories and looks up the original passages when a subtask needs details the gists omit. For base models, we use GPT-5.4~\citep{openai2026gpt54}, Claude Sonnet 4.6~\citep{anthropic2025claude46sonnet}, Gemini 3 Flash~\citep{doshi2025gemini3flash}, and DeepSeek-V4-Pro~\citep{deepseekai2026deepseekv4} for the LongBench-v2 experiments. 

As shown in Table~\ref{tab:multidoc_results}, \toolName{} achieves the highest average accuracy across all four model families: 60.1\% with GPT-5.4, 59.8\% with Claude Sonnet 4.6, 61.5\% with Gemini 3 Flash, and 67.5\% with DeepSeek-V4-Pro. Compared with the strongest baseline for each model family, these correspond to gains of 5.7, 5.3, 4.4, and 3.6 percentage points, respectively. \toolName{} also achieves the best or tied-best result in most domain-model combinations, suggesting that the benefit comes from a general improvement in evidence selection and reuse rather than from a single model or domain.

\subsection{Combining \toolName{} with RLM}\label{sec:with_rlm}
Recursive Language Models (RLMs)~\citep{zhang2025recursive} process long contexts by recursively selecting context segments, issuing sub-calls, and composing the resulting partial answers. Crucially, RLM performs this process through a code-mediated Read-Eval-Print Loop (REPL): the long input is stored as an external variable that the model can inspect, parse, and aggregate programmatically. This makes RLM a natural fit for OOLONG~\citep{bertsch2025oolong}, an aggregation-heavy benchmark where each sample consists of timestamped, user-attributed entries and answering a question often requires classifying many entries before computing a distributional answer.

\begin{table}
\centering
\caption{Comparison of RLM, \toolName{}, and their hybrid on LongBench-v2 Multi-Doc QA and OOLONG using GPT-5.}
\begin{tabular}{lcc}
\toprule
\textbf{Method} & \textbf{Acc.} & \textbf{Cost/Task} \\
\midrule
\multicolumn{3}{l}{\textit{OOLONG}} \\
RLM & 56.0\% & \$0.43 \\
\toolName{} & 53.3\% & \$0.47 \\
\toolName{}+RLM & \textbf{64.0\%} & \textbf{\$0.40} \\
\midrule
\multicolumn{3}{l}{\textit{LongBench-v2 Multi-Doc QA}} \\
RLM & 55.8\% & \$0.29 \\
\toolName{} & 57.9\% & \$0.30 \\
\toolName{}+RLM & \textbf{60.3\%} & \textbf{\$0.24} \\
\bottomrule
\end{tabular}
\label{tab:oolong}
\end{table}

We first evaluate \toolName{} and RLM separately on both LongBench-v2 Multi-Doc QA and OOLONG using GPT-5 with medium reasoning (RLM uses GPT-5-mini for sub-calls and recursion depth=1). As shown in Table~\ref{tab:oolong}, the two methods have complementary strengths. On LongBench-v2, \toolName{} achieves higher average accuracy than RLM at roughly the same cost per task. On OOLONG, however, \toolName{} underperforms RLM in both average accuracy and cost. This is because OOLONG is closer to a structured data-processing benchmark than a conversational reasoning benchmark (such as LongBench-v2): many questions require exact counting, filtering, comparison, and tie handling. In this setting, the natural-language shared context used by \toolName{} is not reliable. RLM performs better because its core interface is code-mediated: through a REPL environment, the model can inspect, parse, transform, and aggregate the input using executable programs and programmatic sub-calls. However, RLM still relies on centralized coordination, where recursive calls report back to a root process that must reconcile and aggregate their outputs. This centralized aggregation is less suitable for natural-language multi-document reasoning, helping explain RLM's weaker LongBench-v2 performance. These results motivate a hybrid design that combines RLM's precise REPL-based execution with \toolName{}'s decentralized coordination.

We therefore combine the two methods by retaining RLM as the underlying reasoner while adding two \toolName{} components: a shared context and a decentralized task queue. In this hybrid setting, there is no main-agent/sub-agent hierarchy; all agents are equally ranked RLM instances that coordinate through shared state. An initial RLM call produces a compact work plan, after which subtasks are placed into the task queue and claimed by parallel RLM workers. Each worker's output is admitted into the shared context, and the system then generates any additional subtasks from the current problem state, and computes the final answer from the shared context once no further subtasks are needed. This design preserves RLM's strength in precise code-mediated execution while adding \toolName{}'s decentralized coordination. As shown in Table~\ref{tab:oolong}, the combined method achieves the best accuracy and lowest cost on both benchmarks, indicating that RLM and \toolName{} are complementary rather than competing approaches.

\section{Why \toolName{} Is Both Accurate and Cost-Efficient on SWE-bench}
\label{sec:swe_analyze}

Three trace-level mechanisms explain why \toolName{} improves both accuracy and cost efficiency. In the \toolName{} traces below, each shared entry is one thread's note, tagged with its thread id (\texttt{t0}, \texttt{t1}, \dots) and a type such as \texttt{FACT}, \texttt{FAIL}, or \texttt{PATCH\_SUMMARY}.

\paragraph{(1) Parallel agents complement each other by sharing failures.}
The first mechanism is that failed hypotheses become reusable state rather than private dead ends. In isolated forks, a negative result remains local to one trajectory, so other attempts may spend their own budget rediscovering the same failure. In \toolName{}, once such a failure is admitted to the shared context, later agents can treat it as a constraint and redirect their search.

This mechanism appears in the trace below. The task fixes \texttt{lambdify} for single-element tuples. The natural guess, the Python printer layer, is a red herring; the real bug is on a separate tuple-building path in \texttt{sympy/\allowbreak utilities/\allowbreak lambdify.py}:

\begin{lstlisting}[style=gist]
[t0/FACT] AbstractPythonCodePrinter change did not affect lambdify output
[t1/FACT] sympy/utilities/lambdify.py:964 _recursive_to_string manual tuple join bypasses printer fix
[t0/FACT] sympy/utilities/lambdify.py:964 manual join for tuples lacks trailing comma logic
\end{lstlisting}
 
The decisive update is the negative result in the first line: thread \texttt{t0} shows that changing the printer does not affect the output. After reading this failure, \texttt{t1} avoids repeating the same detour and localizes the actual bypass in \texttt{\_recursive\_to\_string}; \texttt{t0} then records the concrete defect, the missing trailing comma. Thus, \toolName{} turns a failed hypothesis into shared progress, improving later agents' search efficiency.

\paragraph{(2) Admitted constraints remain binding shared state.}
The second mechanism is that \toolName{} preserves important constraints as
shared state rather than routing them through a central controller. In centralized coordination, a main agent may soften, omit, or reopen constraints discovered by sub-agents before they can guide later work.
 
This mechanism appears in the trace below. Folding multiple search filters into one \texttt{.filter()} call is unsafe for multi-valued relations, where \texttt{.filter(A).\allowbreak filter(B)} and \texttt{.filter(A, B)} differ. In AOrchestra, a sub-agent finds exactly this danger, but the constraint must pass through the main agent, which reopens the optimization and softens it into a tradeoff about reducing joins ``for many use cases''; the run fails. The relevant fact was found, but not preserved as binding state. In contrast, \toolName{} keeps the constraint explicit and reusable:
 
\begin{lstlisting}[style=gist]
[t3/FAIL] django/contrib/admin/options.py:1041 single .filter() breaks M2M multi-term search
[t3/FACT] Django ORM .filter(Q1, Q2) vs .filter(Q1).filter(Q2) differs for multi-valued relations
[t3/FACT] lookup_spawns_duplicates determines if single .filter() is safe
[t1/FACT] lookup_spawns_duplicates preserves M2M search semantics while optimizing FKs
\end{lstlisting}

Thread \texttt{t3} records the unsafe case (\texttt{FAIL}), the semantic reason, and the predicate \texttt{lookup\_spawns\_duplicates}, which determines when the optimization is valid. Thread \texttt{t1} then builds on this state to preserve M2M semantics while still optimizing foreign-key cases. Thus, later threads inherit the binding constraint rather than reopening a globally invalid simplification.

\paragraph{(3) Compact patch summaries carry discoveries.} 
The third mechanism is compact sharing, which lowers cost by preventing peer
communication from becoming another long raw transcript. Sharing full traces
would preserve information, but would also expose every worker to command
history, file dumps, failed edits, and intermediate reasoning. \toolName{}
instead shares a compressed version of the useful discovery, so later workers can reuse the result without reading the entire trajectory.

This mechanism appears in the trace below. Both the shared and no-share settings solve the task, but compact sharing reduces the cost from \$0.399 to \$0.125. The difference is not that \toolName{} shares more context; rather, it shares a compressed version of the useful search result (labeled as \texttt{PATCH\_SUMMARY}). One thread first rules out the ordinary tuple-printer path and localizes the actual bypass in \texttt{lambdify.py}:

\begin{lstlisting}[style=gist]
[t1/FACT] StrPrinter._print_tuple at sympy/printing/str.py:868 already includes trailing comma logic
[t1/FACT] sympy/utilities/lambdify.py:964 lacks trailing comma for single-element tuples
[t1/PATCH_SUMMARY] files=sympy/utilities/lambdify.py | idea=Modify _recursive_to_string to add a trailing comma for single-element tuples | evidence=reproduce_issue.py PASSED
\end{lstlisting}

This patch summary turns a multi-step debugging trajectory into a short,
evidence-backed handoff. Later workers do not need the full command history, file dumps, failed edits, or intermediate reasoning. Thus, compact sharing preserves the reusable discovery while avoiding the cost of exposing every peer to the raw trace, explaining why \toolName{} solves the task at much lower cost.

Together, these examples explain the trend in Table~\ref{tab:swebench_verified}: \toolName{} keeps reusable facts, failures, constraints, and patch summaries visible to all peers, while avoiding both redundant isolated search and the high token cost of raw trace sharing. This improves per-thread success while reducing cost.


\section{Prompts}\label{sec:prompts}

\begin{tcolorbox}[
  enhanced,
  breakable,
  title={DeLM Prompt ($n=2$, Part 1): Initial Instructions},
  colback=gray!3,
  colframe=black!55,
  coltitle=white,
  colbacktitle=black!55,
  fonttitle=\bfseries,
  boxrule=0.5pt,
  arc=2pt,
  left=4pt,
  right=4pt,
  top=4pt,
  bottom=4pt
]
\textcolor{black!70}{\texttt{[DeLM AGENT A]}} \\[2pt]
\textbf{Workspace.} \\
Your task workspace is \{workspace\}. Do not spawn extra agents. \\[3pt]
\textbf{Role.} \\
You are \{worker\}, one of two agents solving this task in parallel. Roster: \{roster\}. \\[3pt]
\textbf{Objective.} \\
Minimize elapsed time to a correct, complete submission. Start useful work promptly, unblock your peer with usable results, and reuse applicable evidence. \\[3pt]
\textbf{Coordination.}
\begin{itemize}[leftmargin=1.4em, itemsep=1pt, topsep=2pt]
  \item Divide work and exchange findings and text files through the board. It is the only connection between workers. Each worker has a separate container, so edits and installs are local. Every worker submits a complete solution from its own workspace.
  \item The board is ready immediately. The first worker creates the initial queue: it reserves its own scope, then publishes actionable peer work or states the concrete dependency preventing a split. After that handoff, both can publish, reserve, and claim distinct work. If the first worker ends before the handoff, the other takes over. Both can share findings immediately.
\end{itemize}

\textbf{Task requirements.} \\
The original task remains authoritative for both workers. Queue items divide work; their summaries never replace or weaken its requirements. Preserve every required observable effect, including state changes and nonvisual side effects. \\[3pt]
\textbf{Performance.}
\begin{itemize}[leftmargin=1.4em, itemsep=1pt, topsep=2pt]
  \item For performance-constrained work, choose the complete first-call path before fixing component interfaces. In an existing initial item or FACT, state what the producer prepares, what the consumer receives directly, and what substantial work remains on the first required invocation. Retained state is not directly usable if that invocation must rebuild another representation. Design preparation and updates within the task's required lifecycle, preserving fresh-start, mutation and lifetime semantics.
  \item For decision or search operations, consider cheap sufficient conditions for exact acceptance and rejection before general preparation or search. An early answer must satisfy every required condition; a heuristic or necessary condition is not enough. Use the complete method when no such condition establishes an answer.
  \item Share the interface early and implement independent parts in parallel. When a specific interface cost could prevent a required target, resolve that concern with existing evidence or a small probe before expanding code dependent on that choice. Keep unaffected work moving. Address a measured bottleneck before extending an expensive comparison; reuse compatible components.
\end{itemize}

\textbf{Your initial work:} \\
Inspect the task and existing files to choose the simplest approach that can satisfy all stated requirements and identify complementary, reusable implementation outputs. Read the board and incorporate useful peer findings. Establish the initial split: reserve your scope and publish actionable, nonoverlapping work for your peer, with available inputs, interfaces and the first usable dependency. Do not wait to complete a whole-task plan before publishing useful work and implementing your part in this same session. If no independent item is clear yet, state your scope and the unresolved dependency so your peer can help resolve it. Add remaining work only as needed.
\end{tcolorbox}

\begin{tcolorbox}[
  enhanced,
  breakable,
  title={DeLM Prompt ($n=2$, Part 2): Board Commands},
  colback=gray!3,
  colframe=black!55,
  coltitle=white,
  colbacktitle=black!55,
  fonttitle=\bfseries,
  boxrule=0.5pt,
  arc=2pt,
  left=4pt,
  right=4pt,
  top=4pt,
  bottom=4pt
]
\textcolor{black!70}{\texttt{[DeLM AGENT A]}} \\[2pt]
\textbf{Board commands:} \\
\textasciigrave{}sh /tools/context.sh ACTION ...\textasciigrave{} through your ordinary shell tool.
\begin{itemize}[leftmargin=1.4em, itemsep=1pt, topsep=2pt]
  \item {\raggedright\texttt{read [-{}-all] [-{}-wait SECONDS] [-{}-for ID]}\par}\nopagebreak[4]
The first view is complete. Later views show active work, new publications and changed statuses since your last served view. Unchanged active items show their identity and owner. Session states (starting/running/ended) are always shown. A session ending does not establish task correctness. Use read -{}-all for older entries or a missed response. A waiting read returns on a peer publication or a peer ending, otherwise after at most 25 seconds. Status changes do not end a wait. -{}-for names a needed item or publication and returns if it is available already.
  \item {\raggedright\texttt{publish -{}-items JSON}\par}\nopagebreak[4]
Publish open items as a list of objects with subject and detail, following the startup roles above. Publishing does not claim ownership. Matching subjects reuse existing items without changing their owner, detail, or completion state. Subject target: 80, detail: 300 characters.
  \item {\raggedright\texttt{claim -{}-item ID}\par}\nopagebreak[4]
First claim wins; a loss names the owner. One unfinished claim per worker.
  \item {\raggedright\texttt{reserve -{}-subject TEXT -{}-detail TEXT}\par}\nopagebreak[4]
Atomically create and claim one scope if your claim slot is free. Matches reuse open items, report claimed owners, or remain done. Subjects match ignoring case and extra whitespace. Different titles can still overlap, so respect visible scopes.
  \item {\raggedright\texttt{done -{}-item ID -{}-summary TEXT [-{}-attach PATH...]}\par}\nopagebreak[4]
Finish your claim with results, affected files/functions, limitations, and check outcomes already obtained (summary target: 800 characters). Attach reusable text files from your workspace or /tmp. Publication returns its immutable ID.
  \item {\raggedright\texttt{note -{}-type FACT|FAIL -{}-text TEXT -{}-scope TEXT -{}-basis TEXT [-{}-attach PATH...]}\par}\nopagebreak[4]
FACT means a finding supported by observations. FAIL means a contradicted approach or hypothesis. State scope and reproducible evidence, including the inputs, conditions and actual results behind measurements. Identify unfinished parts. For measurements, include the input construction, sample count, starting state, metric and aggregation. Attach the runnable command or script when sharing its result. Targets in characters: text 400, scope 120, basis 200. Keep attachments focused (target: 20 files and 200,000 bytes per bundle).
  \item {\raggedright\texttt{apply -{}-id ID [-{}-only PATH...]}\par}\nopagebreak[4]
Copy attachments over original or registered peer versions. Unknown local edits refuse the selected batch; inspect and reconcile locally. Identical files and newer peer versions are kept.
  \item {\raggedright\texttt{expand -{}-id ID [-{}-only PATH...]}\par}\nopagebreak[4]
Show attachment diffs against task originals, not your current files. New files and -{}-only paths are shown in full.
  \item {\raggedright\texttt{status -{}-text TEXT}\par}\nopagebreak[4]
State a changed question, configuration, dependency or scope (target: 200 characters). Claim, reserve and done already set work status.
\end{itemize}
\end{tcolorbox}

\begin{tcolorbox}[
  enhanced,
  breakable,
  title={DeLM Prompt ($n=2$, Part 3): Coordination},
  colback=gray!3,
  colframe=black!55,
  coltitle=white,
  colbacktitle=black!55,
  fonttitle=\bfseries,
  boxrule=0.5pt,
  arc=2pt,
  left=4pt,
  right=4pt,
  top=4pt,
  bottom=4pt
]
\textcolor{black!70}{\texttt{[DeLM AGENT A]}} \\[2pt]
\textbf{Reading the board.} \\
Read the board at startup, before a new implementation subproblem or substantial diagnostic hypothesis, before final integration, and before submitting. Choose a new diagnostic after receiving the view, not in the invocation that first requests it. If a peer owns or answered that question, choose a distinct unresolved one unless changed conditions or contradictory evidence justify repetition. Command responses already include fresh board views, which replace a separate read at those boundaries. Routine local checks do not each require a read, status, or reservation. \\[3pt]
\textbf{Sharing results.} \\
Share usable changes promptly when they can unblock a peer. State unfinished parts, known limitations, interfaces, and existing check evidence. New checks are not required merely to share partial work or close a subtask. Closing a subtask does not establish that the complete solution meets the requirements. Do not hold reusable work until unrelated work is finished. When outputs or service state cannot be attached, share the generating or replay code and exact command and inputs to reproduce them locally. When no decision depends on a returned view, combine a board publication and the next local command in one shell call. Avoid repeating evidence across board fields and the final answer. Do not add documentation or a new build wrapper unless the task requires it or it is needed to produce or check the submitted output. \\[3pt]
\textbf{Waiting:} \\
do useful independent implementation, integration or diagnosis when available. Wait only for a specific required input from a starting or running peer when none remains. Name the missing input in status and inspect the returned board view before deciding to wait. Do not batch that response with a blocking wait that hides it. If still needed, use read -{}-wait 25 (-{}-for ID when known) with yield\_time\_ms: 30000, reassessing each returned view before waiting again. Never use sleep, repeated polling, duplicate probes, redundant checks or paperwork to fill the gap. Never wait for an ended peer or a nonrequired addition. \\[3pt]
\textbf{Long commands.} \\
Run necessary long diagnostics asynchronously when useful independent work remains, exposing intermediate results. Preserve their inputs and code; results apply to those versions. Read new output when it can change a decision instead of repeatedly polling. If a peer update could make a tool wait unnecessary, read the board before blocking, not after the wait in the same call. Avoid competing workload during timing measurements. Wait only for evidence still required for completion. \\[3pt]
\textbf{The queue:}
\begin{itemize}[leftmargin=1.4em, itemsep=1pt, topsep=2pt]
  \item Queue items are for required implementation, repair of a demonstrated defect, or a concrete unanswered question needed to implement the solution. Do not create separate verification, testing, review, reproduction, or solution-checker assignments, including at a peer's request. This rule applies to queue items created with publish or reserve. Diagnosis and measurements needed to choose or repair an implementation remain allowed. A checker that is itself the requested product is implementation.
  \item Name each item's output, available inputs, shared interface and first usable dependency. Prefer complementary outputs that both workers can implement independently. Preserve required public behavior, interfaces and side effects. Reuse compatible components. Pursue a different approach to resolve a concrete limitation, sharing compatible parts. Keep related shared-file edits together. Do not invent work just to occupy a peer.
  \item After the initial handoff, claim an existing item when it fits; otherwise reserve a distinct required implementation scope. Respect the returned owner. Ownership covers an output, not a fixed algorithm. Publish a needed component when a peer can implement it independently, stating its interface, inputs and nonoverlapping scope while continuing your part. Keep small local adjustments under the current claim. Routine local integration and necessary checking need no separate item or active claim. Respect ownership and interfaces while the owner is running. Report concrete defects through scoped FAIL notes. To hand off incomplete work, close the claim with its remaining scope and reusable files. An ended worker's claim stays with it, but cannot block you from completing required work. Keep completed items closed. Do not assign verification roles through notes or status. Requests for existing evidence and interface clarification remain useful. Initial planning does not make either worker a permanent coordinator or checker. There is no global plan-closure step.
\end{itemize}
\end{tcolorbox}

\begin{tcolorbox}[
  enhanced,
  breakable,
  title={DeLM Prompt ($n=2$, Part 4): Validation and Completion},
  colback=gray!3,
  colframe=black!55,
  coltitle=white,
  colbacktitle=black!55,
  fonttitle=\bfseries,
  boxrule=0.5pt,
  arc=2pt,
  left=4pt,
  right=4pt,
  top=4pt,
  bottom=4pt
]
\textcolor{black!70}{\texttt{[DeLM AGENT A]}} \\[2pt]
\textbf{Validation and completion:}
\begin{itemize}[leftmargin=1.4em, itemsep=1pt, topsep=2pt]
  \item Derive necessary checks from the original task, not the queue summary or the current implementation. Preserve its measurement domain and apply each stated exception only to the requirement it modifies. Assess your own and peer evidence against those requirements. Reuse applicable results; adoption alone does not require repeating a check. Evidence must match the relevant code, inputs, configuration and environment of the final local output.
  \item For a new comparison, obtain matched candidate and reference observations before collecting repeats. Repeat as required by the task or where uncertainty could change the implementation or acceptance decision, preserving the required starting conditions. Preserve inputs, expected results and reference measurements. After candidate-only changes, rerun affected candidate checks; reuse reference measurements only under matching inputs, reference version, procedure and operating conditions.
  \item Publish an observed relevant failure promptly as FAIL with its conditions. Keep it unresolved until corrected code or a task-based explanation addresses it; changing the sample or averaging it with unrelated cases is not a repair. When ground truth is unavailable, use available references, convergence or invariants and state what remains unmeasured. Never turn that limitation into a claim of measured success.
\end{itemize}

\textbf{Submission.} \\
Before submitting, read the board and address published corrections relevant to your final local solution. A required repair is a dependency even if an earlier artifact is usable. Do not finish with a known unresolved failure of a required behavior or performance target. Once applicable evidence supports satisfying all requirements and no relevant failure remains, preserve that candidate and submit. An optional stronger result or an already-running search is not a reason to delay. Prevent exploratory work from changing the submitted output. Close any claim you still hold with done. If your item is already done, publish required corrections in a note rather than opening a claim merely to finish it. Submit a concise answer with required explanations and known limitations. A peer still working does not block a complete local submission.
\end{tcolorbox}

\paragraph{Four-agent configuration.}
For $n=4$, the two additional agents initially identify reusable components and interfaces, and resolve a concrete design uncertainty. These directions yield to available implementation work. The prompt also clarifies cross-author file conflicts, interface changes, local state and artifact reuse, container resource limits, and measurement intervals. Shared instructions and the first two startup prompts also contain wording changes.

\end{document}